\documentclass[utf8]{frontiersFPHY} 

\setcitestyle{square} 
\usepackage{url,hyperref,lineno,microtype}
\usepackage[onehalfspacing]{setspace}

\usepackage{amssymb}
\usepackage{amsmath}
\usepackage{amsfonts}
\usepackage{epsfig}
\usepackage{bbm}
\usepackage{comment}
\usepackage{multirow}
\usepackage{longtable}
\usepackage{array}
\usepackage{graphics}
\usepackage{tikz}
\usetikzlibrary{matrix}

\def\keyFont{\fontsize{8}{11}\helveticabold }
\def\firstAuthorLast{V. Som\`a}
\def\Authors{Vittorio Som\`a$^\ast$}
\def\Address{IRFU, CEA, Universit\'e Paris-Saclay, 91191 Gif-sur-Yvette, France 
 }
\def\corrAuthor{Vittorio Som\`a}
\def\corrEmail{vittorio.soma@cea.fr}

\newcommand{\ad}[1]{a_{#1}}
\newcommand{\ac}[1]{a^{\dagger}_{#1}}
\newcommand{\bra}[1]{\langle #1 \vert}
\newcommand{\ket}[1]{\vert #1 \rangle}
\newcommand{\ii}{\textrm{i}}

\def\a{\alpha}
\def\b{\beta}

\def\d{\delta}
\def\g{\gamma}

\def\tt{\hat \tau}

\def\om{\omega}

\def\ran{\rangle}

\def\lan{\langle}
\def\ll{\langle\!\langle}

\def\ema{$N\!N\!$+$3N(400)$}
\def\lnl{$N\!N\!$+$3N\text{(lnl)}$}
\def\sat{NNLO$_{\text{sat}}$}

\begin{document}
\onecolumn
\firstpage{1}

\title{Self-consistent Green's function theory for atomic nuclei} 

\author[\firstAuthorLast ]{\Authors} 
\address{} 
\correspondance{} 

\extraAuth{}

\maketitle

\begin{abstract}
Nuclear structure theory has recently gone through a major renewal with the development of \textit{ab initio} techniques that can be applied to a large number of atomic nuclei, well beyond the light sector that had been traditionally targeted in the past.
Self-consistent Green's function theory is one among these techniques. 
The present work aims to give an overview of the self-consistent Green's function approach for atomic nuclei, including examples of recent applications and a discussion on the perspectives for extending the method to nuclear reactions, doubly open-shell systems and heavy nuclei.

\tiny
 \keyFont{ \section{Keywords:} nuclear theory, many-body theory, \emph{ab initio} nuclear structure, Green's function theory, open-shell nuclei}
\end{abstract}

\vskip0.2cm
\section{Introduction}
\vskip0.2cm

The theoretical description of atomic nuclei is particularly challenging, for several reasons.
First, different energy scales are at play, which is manifest in the rich set of observables and phenomena one is confronted with\footnote{One could further recall the large variety of experimental probes used to study nuclear properties, as well as the diversity of processes that can concern atomic nuclei, for which three of the four fundamental interactions (strong, electromagnetic and weak) are involved.}.
As a consequence, the choice of relevant degrees of freedom might not be universal but depend on the particular properties one is interested in.
The standard description in terms of nucleons, i.e. protons and neutrons, leads to a many-body Schr\"odinger equation for up to a few hundred particles, which are too many to be easily treated exactly but too few to be amenable to a statistical treatment\footnote{This characteristic is usually associated to the definition of nuclei as \textit{mesoscopic} systems.}.
Furthermore, interactions between nucleons should be derived from quantum chromodynamics (QCD) in its nonperturbative regime, which prevents direct calculations and requires an additional modelisation effort\footnote{Recently, pioneering lattice QCD calculations have produced the first usable nucleon-nucleon potentials~\cite{Ishii07a, McIlroy18}. However, complications still exist in the derivation of three-nucleon forces such that the systematic implementation of lattice QCD interactions is not envisaged in the very near future.}.
In the past, all these features have hindered full solutions of the many-body Schr\"odinger equation and favoured the development of a plentitude of models, following different strategies and characterised by different ranges of application.
Although successful in reproducing experimental observations, these models are not always comparable to each other and do not provide a coherent and unified description of nuclear systems.

Only relatively recently so-called \textit{ab initio} calculations of atomic nuclei, i.e. systematically improvable solutions of the many-body Schr\"odinger equation that start solely from the knowledge of inter-nucleon interactions, have become available for a large number of isotopes.
These advances were made possible by the concomitance of different factors.
First, new formal developments of many-body techniques were carried out.
Second, chiral effective field theory ($\chi$-EFT) was introduced in nuclear physics~\cite{Epelbaum09, Machleidt11, Hammer20}, providing a systematic and consistent framework in which the nuclear Hamiltonian can be modelled.
Third, similarity renormalisation group (SRG) techniques were applied to $\chi$-EFT Hamiltonians~\cite{Bogner10}, which largely improves the convergence of actual calculations with respect to basis truncations, thus enabling the handling of heavier nuclei.
Finally, the availability of computational tools and in particular high-performance computing resources largely increased.
As a result, at present, \textit{ab initio} calculations (with two- and three-nucleon interactions) can be performed for a considerable fraction of the nuclei with mass number $A \lesssim 100$.

Among the techniques applicable to the nuclear many-body problem, one can distinguish ``virtually exact'' and ``approximate'' approaches.
Approaches in the first category do not impose any \textit{formal} approximation on the solution of the many-body Schr\"odinger equation, which is thus affected only by basis truncation and numerical errors.
Typical examples of such methods are Quantum Monte Carlo~\cite{Carlson15, Gandolfi20}, no-core shell model (or configuration interaction)~\cite{Barrett13} or nuclear lattice EFT~\cite{Lahde14}.
The second category includes techniques that do approximate the solution of the Schr\"odinger equation in a systematic way.
This solution can be improved, in principle, up to the recovery of the exact one.
This is typically achieved by first selecting a reference state and then defining an expansion on top of it, usually in terms of particle-hole correlations (whence the denomination \textit{correlation-expansion} methods).
In doing so, the main advantage resides in the scaling with the basis size:
while exact methods scale factorially or exponentially with the system size, correlation-expansion methods only scale polynomially and can be thus applied to a much wider set of nuclei.

The simplest (and most inexpensive) among correlation-expansion approaches is many-body perturbation theory (MBPT)~\cite{shavitt09a}.
Following the belief that the nuclear many-body problem is intrinsically non-perturbative, MBPT was put aside for several decades after its development in the 1950s.
Only relatively recently, with the advent of EFTs and, specially, SRG techniques applied to nuclear Hamiltonians, MBPT was revived~\cite{Tichai16, Tichai18} and it is now considered as a convenient approach for large-scale systematic calculations.
Moreover, it can be easily complemented with resummation techniques like Pad\'e or eigenvector continuation~\cite{Demol20a, Demol20b} (see Ref.~\cite{Tichai20} for a recent review).
Other methods rely on the MBPT concept but are built such that infinite subsets of MBPT contributions are resummed by construction. 
The three typical examples employed in the nuclear context are in-medium similarity renormalisation group (IMSRG), coupled-cluster (CC) and self-consistent Green's function (SCGF) methods.
While IMSRG has been originally designed for the nuclear many-body problem not long ago~\cite{Tsukiyama11, Hergert16}, CC and SCGF, although initially proposed for nuclear systems, have been largely developed in quantum chemistry and solid-state physics before being reimported into nuclear physics starting from the 1990s~\cite{Dickhoff92, Kowalski04} (see Ref.~\cite{Hagen14} for a recent CC review).
IMSRG~\cite{Bogner14}, CC~\cite{Jansen14}, as well as no-core shell model~\cite{Smirnova19} have also been adapted to derive a valence-space interaction that can be used in standard shell-model codes, thus further enlarging the reach of \textit{ab initio} calculations.

The present manuscript deals with one of such correlation-expansion approaches, the self-consistent Green's function method.
Although GFs have been and are implemented in different ways in the context of atomic nuclei (see e.g.~\cite{Broglia16}), here the focus is on \textit{ab initio} SCGFs.
Early SCGF calculations with realistic nucleon-nucleon potentials dealt mainly with infinite nuclear matter~\cite{Dickhoff92}.
Formal developments setting the bases for modern implementations in finite nuclei date back to the early 2000s~\cite{Barbieri01, Barbieri02, Dickhoff04}.
By the end of the decade advanced SCGF implementations were routinely applied~\cite{Barbieri09a, Barbieri09b}.
In 2011, standard (i.e., Dyson) SCGF theory was generalised to a particle-number-breaking (i.e., Gorkov) framework~\cite{Soma11, Soma14a}.
In 2013 the inclusion of three-body interactions was formalised in Dyson theory~\cite{Carbone13}.
In 2018, working equations for the state-of-the-art many-body truncation used in nuclear structure calculations (algebraic diagrammatic construction at third order, see Sec.~\ref{sec_ADC}) were derived~\cite{Raimondi18}.

Building on these formal advances, several applications (based on either Dyson or Gorkov frameworks) have been carried out in the past decade.
Typical examples are ground-state properties of medium-mass nuclei, going from the oxygen region~\cite{Cipollone13, Cipollone15} to silicon and sulfur~\cite{Duguet17}, calcium~\cite{Soma13, Soma14b} and nickel~\cite{Soma20} regions, up to the very recent computation of tin and xenon isotopes~\cite{Arthuis20}.
As discussed in Secs.~\ref{sec_basics} and \ref{sec_excited}, excited spectra of odd-even nuclei are also easily accessible in SCGF theory and were studied in particular in Refs.~\cite{Cipollone15, Duguet17, Soma20}.
Excited states of even-even systems were addressed in the form of the electromagnetic dipole response in Ref.~\cite{Raimondi19a}.
In Ref.~\cite{Idini19}, \textit{ab initio} optical potentials were computed and applied in elastic scattering off $^{16}$O and $^{40}$Ca.
Ref~\cite{Barbieri19} instead discussed lepton-nucleus scattering, with particular attention to neutrino scattering off $^{40}$Ar. 
Dedicated applications studied effective charges (typically employed in shell-model calculations)~\cite{Raimondi19b} and the scale dependence of effective single-particle energies and other non-observable quantities~\cite{Duguet15}.
The possibility of using nucleon-nucleon interactions derived from lattice QCD calculations was also explored in Ref.~\cite{McIlroy18}.
Last but not least, several collaborations with experimental groups have led to testing the method on e.g. state-of-the-art measurements of nuclear masses~\cite{Rosenbusch15, Leistenschneider18}, energies and spins of excited states~\cite{Flavigny13, Papuga14, Chen19, Sun20}, charge and matter radii~\cite{Lapoux16}.

The present article discusses a few of these examples, without any pretension of exhaustivity but with the aim of giving the reader a perception of the reach and versatility of SCGF method, as well we the great deal of possible applications to atomic nuclei.
A short introduction to the formalism and implementation in finite nuclei is presented beforehand; however, again, it is far from being complete.
The reader interested in a comprehensive treatment of GF theory in a modern form is referred e.g. to the book of Dickhoff and Van Neck~\cite{DickhoffVanNeck}.
Older, yet valuable sources are the books of Nozi\`eres~\cite{nozieres64} and Abrikosov, Gorkov and Dzyaloshinski~\cite{AbrikosovBook}.
An interesting work, although tailored to solid-state physics, is the one of Economou~\cite{EconomouBook}.
An extensive review covering SCGF applications to nuclear physics appeared in 2004~\cite{Dickhoff04}.
A more recent pedagogical introduction to the basics of SCGF formalism and implementation in both finite nuclei and infinite nuclear matter (including computational details and examples of numerical codes) can be found in Ref.~\cite{Barbieri17}. 
A numerical code including a second-order evaluation of the self-energy in the Dyson framework is publicly available~\cite{BoccaDorataPublic}.

This manuscript is organised as follows.
In Sec.~\ref{sec_basics} the most important concepts and equations of GF theory are introduced. 
Section~\ref{sec_implementation} describes the actual implementation of the methods in calculations of finite nuclei, briefly reviewing the most commonly used self-energy truncations, the working equations, the choice of basis and interaction.
In Sec.~\ref{sec_applications} some representative applications to ground-state properties, excitation spectra and lepton-nucleus collisions are discussed.
Finally, section~\ref{sec_perspectives} addresses perspectives and present as well as future challenges, focusing on three directions of research: the consistent treatment of nuclear structure and reactions, the generalisation to doubly open-shell systems and the extension to heavy nuclei.

\vskip0.2cm
\section{Basic concepts and equations}
\label{sec_basics}
\vskip0.2cm

Many-body Green's function\footnote{Let us remark that \textit{many-body Green's functions} and \textit{Green's function Monte Carlo} are \textit{not} the same thing. The latter refers to a (virtually exact) technique that aims at projecting out the ground-state wave function typically from a variational solution of the Schr\"odinger equation~\cite{Carlson15, Gandolfi20}.} theory comprises a set of techniques that originated in quantum field theory (QFT) and have been subsequently imported in the (non-relativistic) quantum many-body problem. 
The late 1950's and the 1960's marked the beginning of the field, with flow of QFT ideas and development of formalism.
Since the 1970's technical developments were realised and the approach was applied throughout several disciplines and types of many-body problems, ranging from many-electron systems such as crystals, molecules and atoms to many-nucleon systems such as nuclei and nuclear matter.
Starting from the 1990's such techniques were implemented as an \textit{ab initio} 
method in nuclear physics.

As for other many-body methods, the purpose of such techniques is to achieve an approximate (yet systematically improvable) solution of the $A$-body Schr\"odinger equation.
In standard (or \textit{Dyson}) many-body Green's function theory, this is realised by rewriting the Schr\"odinger equation in terms of one-, two-, ..., $A$-body objects $g^{I}(=g), g^{II}, ..., g^{A}$ named propagators or, indeed, Green's functions (GFs).
Each of these objects is then expanded in a perturbation series, which in practical applications is truncated to include a subset of all possible contributions.
In \textit{self-consistent} schemes such series are themselves expressed in terms of the exact GFs, which requires an iterative solution and makes the method intrinsically non-perturbative, effectively resumming an infinite subsets of perturbative terms.
The $x$-body GF $g^{x}$ allows one to compute all $x$-body observables in the $A$-body ground state\footnote{One notable exception is the ground-state energy $E_0^A$: for a Hamiltonian containing up to $y$-body operators the knowledge of one- up to $(y-1)$-body Green's functions is sufficient,
see Eqs.~\eqref{eq_Koltun} and ~\eqref{eq_Koltun_W}.
}.
Therefore, for most applications one is mainly interested in the one-body GF.

Formally $g$ is defined as the expectation value of a time-ordered product of annihilation and creation operators in the $A$-body ground state $\ket{\Psi_0^A}$
\begin{equation}
g_{\a\b} (t_\a, t_\b) \equiv -i~\bra{\Psi_0^A} \mathcal{T} [ a_{\alpha}(t_\a) a^{\dagger}_{\b}(t_\b) ] \ket{\Psi_0^A} \: ,
\label{eq_GF}
\end{equation}
with Greek indices labelling basis states of the one-body Hilbert space ${\mathcal H}_1$.
One usually works with the Fourier transform of~\eqref{eq_GF} in the energy domain and recasts its perturbation series into the Dyson equation
\begin{equation}
  g_{\a\b}(\om)=g^{(0)}_{\a\b}(\om)+ \sum_{\g\d}g^{(0)}_{\a\g}(\om) \, \Sigma_{\g\d}^{\star}(\om) \, g_{\d\b}(\om)  \; ,
\label{eq_Dyson}
\end{equation}
where $g^{(0)}$ represents some initial ansatz for $g$, e.g. stemming from the solution of Hartree-Fock (HF) equations.
The (irreducible) self-energy $\Sigma^{\star}$ encodes all terms of the expansion, which is truncated in actual calculations.

Once the one-body GF is obtained as the solution of Eq.~\eqref{eq_Dyson}, the expectation value of any one-body operator ${\hat O}^{1B} \equiv \sum_{\a \b}  O^{1B}_{\a \b} a^{\dagger}_{\a}a_{\b} $ can be computed as
\begin{equation}
\langle {\hat O}^{1B}\rangle  
= \sum_{\a\b} \int_{C \uparrow} \frac{d\om}{2\pi i} \, O^{1B}_{\a \b}~g_{\b\a}(\om)
=\sum_{\a\b}  O^{1B}_{\a \b}\,\,\rho_{\b\a} \; ,
\label{eq_1Bop}
\end{equation}
where $C$$\uparrow$ denotes an integral closed on the upper imaginary plane and the one-body density matrix
\begin{equation}
  \rho_{\alpha \beta} \equiv \bra{\Psi_0^A}a^{\dagger}_{\b}a_{\alpha} \ket{\Psi_0^A}
  = \int_{C \uparrow} \frac{d\om}{2\pi i} \, g_{\a\b}(\om)
\label{eq_DM}
\end{equation}
has been introduced.

Additionally, for a Hamiltonian with one- and two-body operators, the one-body propagator gives access to the total energy by means of the Galitski-Migdal-Koltun (GMK) sum-rule~\cite{Galitskii58, Koltun72}
\begin{equation}
E^A_0 = \sum_{\alpha\beta} \frac{1}{2} 
\int_{C \uparrow} \frac{d\om}{2\pi i} \, [\,T_{\alpha\beta}+\omega\,\delta_{\alpha\beta}\, ]
\, g_{\b\a}(\om) \, ,
\label{eq_Koltun}
\end{equation}
where $T_{\a\b}$ denote the matrix elements of the one-body operator.
Nowadays, realistic nuclear structure calculations require the inclusion of at least a three-body interaction in the starting Hamiltonian.
In this case, the GMK sum rule needs to be generalised to~\cite{Carbone13}
\begin{equation}
E^A_0 = \sum_{\alpha\beta} \frac{1}{2} 
\int_{C \uparrow} \frac{d\om}{2\pi i} \, [\,T_{\alpha\beta}+\omega\,\delta_{\alpha\beta}\, ]
\, g_{\b\a}(\om)
-  \frac{1}{2} \langle W\rangle \, ,
\label{eq_Koltun_W}
\end{equation}
where the ground-state expectation value of the three-nucleon operator $\hat{W}$ has to be evaluated.
Such a term requires in principle the knowledge of the three-body propagator $g^{III}$. 
This is however currently out of reach and in most of practical applications the last term in Eq.~\eqref{eq_Koltun_W} is computed as
\begin{equation}
    \lan W\ran\simeq\frac{1}{6} \, \sum_{\a\b\mu\g\d\nu} W_{\a\b\mu,\g\d\nu}~\rho_{\g\a}~\rho_{\d\b}~\rho_{\nu\mu} \; ,
   \label{eq_Wddd}
\end{equation}
i.e. by approximating the full three-body density matrix with the antisymmetrised product of the one-body one. 
In Ref.~\cite{Cipollone13} this was shown to introduce errors smaller than 250 keV for the binding energy of oxygen isotopes.

In addition to giving access to the ground-state properties of the $A$-body system, the one-body GF contains information on neighbouring ($A\pm1$) nuclei. 
It becomes evident when rewriting the propagator~\eqref{eq_GF} in the Lehmann representation
\begin{align}
 g_{\alpha \beta}(\omega) ~=~& 
 \sum_n  \frac{ 
          \bra{\Psi^A_0}  	\ad{\a}   \ket{\Psi^{A+1}_n}
          \bra{\Psi^{A+1}_n}  \ac{\b}  \ket{\Psi^A_0}
              }{\omega - E_n^{+} + \ii \eta }  \nonumber\\
 ~+~ &\sum_k \frac{
          \bra{\Psi^A_0}      \ac{\b}    \ket{\Psi^{A-1}_k}
          \bra{\Psi^{A-1}_k}  a_{\a}	    \ket{\Psi^A_0}
             }{\omega - E_k^{-} - \ii \eta } \; ,
\label{eq_Lehmann}
\end{align}
where $\ket{\Psi^{A\pm1}_i}$ represent eigenstates of ($A\pm1$)-body systems while 
\hbox{$E_n^{+}\equiv(E^{A+1}_n - E^A_0)$} and 
\hbox{$E_k^{-}\equiv(E^A_0 - E^{A-1}_k)$}
are one-nucleon addition and removal energies, respectively.

In Dyson GF theory the expansion starts from a particle-number conserving (e.g. a Hartree-Fock) reference state providing $g^{(0)}$.  
On top of this, spherical symmetry is typically imposed.
While such an expansion can suitably address closed-shell systems, it becomes inefficient or even breaks down as soon as pairing and/or quadrupole correlations become important. 
If one wishes to stick with a single-reference method, a possible solution consists in working, from the outset, with a symmetry-breaking reference state. 
In particular, breaking U(1) symmetry associated with particle number conservation\footnote{In the case of atomic nuclei proton and neutron numbers are conserved individually, therefore it is always intended U(1)$_N \otimes $U(1)$_Z$ where one of the two or both are broken.} while maintaining spherical symmetry allows one to efficiently capture pairing correlations, thus gaining access to singly open-shell nuclei. 

Dyson GFs were thus generalised to a U(1) symmetry-breaking (typically Hartree-Fock-Bogolyubov) reference state originally by Gorkov~\cite{gorkov}.
The formalism was then adapted and implemented for applications to finite nuclei in Ref.~\cite{Soma11}.
Technically, the extension is achieved by working with an $A$-body ground state that is a linear combination of states with different particle numbers
\begin{equation}
\Psi_0^A \longrightarrow  \Psi_0 = \sum_{A'} c_{A'} \Psi_0^{A'} \, .
\end{equation}
This leads to the definition of four one-body propagators
\begin{subequations}
\label{eq:gg}
\begin{equation}
\label{eq:gg11}
g^{11}_{\a\b}(t,t') \equiv -i \,
\langle \Psi_0 |  \mathcal{T} [
a_{\a}(t) a_{\b}^{\dagger}(t') 
]
| \Psi_0 \rangle \: ,
\end{equation}
\begin{equation}
\label{eq:gg12}
g^{12}_{\a\b}(t,t') \equiv -i \,
\langle \Psi_0  | \mathcal{T} [
a_{\a}(t) \bar{a}_{\b}(t')
]
| \Psi_0 \rangle \: ,
\end{equation}
\begin{equation}
\label{eq:gg21}
g^{21}_{\a\b}(t,t') \equiv -i \,
\langle \Psi_0  | \mathcal{T} [
\bar{a}_{\a}^{\dagger}(t) a_{\b}^{\dagger}(t')
]
| \Psi_0 \rangle \: ,
\end{equation}
\begin{equation}
\label{eq:gg22}
g^{22}_{\a\b}(t,t') \equiv -i \,
\langle \Psi_0  | \mathcal{T} [
\bar{a}_{\a}^{\dagger}(t) \bar{a}_{\b}(t')
]
| \Psi_0 \rangle \: ,
\end{equation}
\end{subequations}
two of which ($g^{11}$ and $g^{22}$) involve \emph{normal} combinations of $a$ and $a^\dagger$ and are associated to the standard density matrix~\eqref{eq_1Bop}.
The remaining two propagators ($g^{12}$ and $g^{21}$) invoke so-called \emph{anomalous} contributions of $a$ and $a^\dagger$ (interpreted as the annihilation or the creation of a nucleon pair) and lead to the definition of an anomalous (or pairing) density matrix
\begin{equation}
  \tilde{\rho}_{\alpha \beta} \equiv \bra{\Psi_0}\bar{a}_{\b}a_{\alpha} \ket{\Psi_0}
  = \int_{C \uparrow} \frac{d\om}{2\pi i} \, g^{12}_{\a\b}(\om) \: .
\label{eq_DMan}
\end{equation}
In Eqs.~\eqref{eq:gg} creation operators $\{\bar{a}_\a^{\dagger} \}$ define a one-body basis dual to $\{a_\a^{\dagger} \}$ and are obtained via
\begin{equation}
\label{eq:gen_aad}
\bar{a}_{a}^{\dagger}(t) \equiv \eta_a a_{\tilde{a}}^{\dagger}(t)\, , \qquad
\bar{a}_{a}(t) \equiv \eta_a a_{\tilde{a}}(t) \: ,
\end{equation}
which correspond to exchanging the state $a$ with its time-reversal partner $\tilde{a}$ up to the phase $\eta_a$~\cite{Soma11}. 
The four Gorkov propagators~\eqref{eq:gg} can be conveniently recast in a $2 \times 2$ matrix notation via Nambu's formalism~\cite{Nambu60}
\begin{equation}
\mathbf{g}_{\a\b} (t,t') \equiv
\left(
\begin{tabular}{cc}
$g^{11}_{\a\b}(t,t')$ & $g^{12}_{\a\b}(t,t')$ \\
& \\
$g^{21}_{\a\b}(t,t')$ & $g^{22}_{\a\b}(t,t')$
\end{tabular}
\right) \: .
\end{equation}
All quantities (operators, self-energy, ...) can be generalised in an analogous fashion such that one ends up with the Gorkov equation
\begin{equation}
\mathbf{g}_{\a\b}(\om)=\mathbf{g}^{(0)}_{\a\b}(\om)+ \sum_{\g\d} \mathbf{g}^{(0)}_{\a\g}(\om) \, \mathbf{\Sigma}_{\g\d}^{\star}(\om) \, \mathbf{g}_{\d\b}(\om)  \; .
\label{eq_Gorkov}
\end{equation}
Similarly, all standard GF equations including Eqs.~\eqref{eq_1Bop}-\eqref{eq_Lehmann} are rewritten in a matrix form.
Last but not least, a chemical potential $\lambda$ needs to be introduced to guarantee that the number of particles is the correct one \textit{on average}. 
This amounts to replacing the Hamiltonian $\hat{H}$ with the grand potential 
\begin{equation}
\hat{\Omega} \equiv \hat{H} - \lambda \hat{A} \, .
\label{eq_grand}
\end{equation}
As a consequence of the symmetry breaking, observables might be contaminated by components associated to different particle numbers.
Even if in practice the variance is expected to remain small\footnote{Away from closed-shell systems, contributions from components with $A'\neq A$ are assumed to cancel out to some extent.
The largest contamination is expected in differential observables across closed shells, where one of the two systems does not spontaneously break particle-number symmetry.}, the broken symmetry has to be eventually restored. While symmetry-restored formalism has been developed for other (post-Hartree-Fock-Bogolyubov) many-body methods~\cite{Duguet:2014jja, Duguet17b}, it remains to be formulated for Gorkov GFs.

\vskip0.2cm
\section{Implementation for atomic nuclei}
\label{sec_implementation}
\vskip0.2cm

\vskip0.2cm
\subsection{Choice of approximation scheme}
\label{sec_ADC}
\vskip0.2cm

Self-consistent GF approximation schemes are defined by the content of the irreducible self-energy, which is expressed as a function of the exact GFs and encodes its perturbative expansion.
There exist several ways of approximating the self-energy.
The most basic one simply amounts to truncating the perturbative expansion at a certain order.
More refined techniques resort to including infinite subsets of perturbation-theory terms via the definition of implicit equations.
This is the case, e.g., of the so called \emph{ladder} or \emph{in-medium T-matrix} approximation that resums all multiple particle-particle scattering contributions\footnote{Other types of resummation are employed in other domains.
For instance, in solid-state physics the resummation of particle-hole (i.e., ring) diagrams (typically in the so-called $GW$ approximation) allows resolving the long-range features of the Coulomb force~\cite{Onida02}.}.
This approximation scheme gained considerable attention in early \emph{ab initio} applications because of its ability of tackling nucleon-nucleon interactions with strong short-range components~\cite{Bozek99, Frick03, Soma06, Rios06, Rios08, Soma08}.
These truncations and resummations are typically conveniently expressed in terms of Feynman diagrams, which facilitate the manipulation of the various terms and give an insight in their physical content.

An alternative, although not orthogonal, route was proposed in the context of quantum chemistry~\cite{Schirmer82, Schirmer83} and instead exploits the analytical structure of the self-energy. 
Similarly to the one-body GF, the exact (dynamical, i.e. energy-dependent, part of the) self-energy displays a Lehmann representation
\begin{equation}
\Sigma_{\a\b}^{\star}(\omega) = 
\sum_{nn'} M^\dagger_{\a n} \left [ \frac{1}{\omega - (E^>+C) + i \eta} \right ]_{nn'} M_{n' \b} +
\sum_{kk'} N^\dagger_{\a k} \left [ \frac{1}{\omega - (E^<+D) - i \eta} \right ]_{kk'} N^\dagger_{k' \b} \; ,
\label{eq_sigma_Lehmann}
\end{equation}
where the matrices $M, N$ couple the single-particle motion of the nucleons (i.e. the one-body propagator in which the self-energy is inserted) to intermediate multiparticle-multihole configurations, whose energies (``bare'' and resulting from the interference between them) are encoded in the matrices $E^>, E^<$ and $C, D$ (respectively).
The algebraic diagrammatic construction at order $n$ [ADC(n)] is built by demanding that, in addition to including all perturbation-theory contributions up to a given order $n$, the approximated self-energy has the same analytical structure as the exact one, i.e., in particular, is the same function of the energy.
The latter condition requires the self-energy to contain additional sets of contributions, e.g. infinite resummations that would necessitate ad-hoc procedures are in this way automatically included in the ADC formalism.
The first order, ADC(1), is simply the standard Hartree-Fock (or Hartree-Fock-Bogolyubov, in the case of Gorkov GFs) approximation. 
ADC(2) introduces lowest-order dynamical correlations in terms of two particle-one hole and two hole-one particle contributions.
ADC(3) builds couplings between such configurations and, as a result, includes infinite-order resummations of both particle-particle/hole-hole and particle-hole ladders.
Higher orders build on higher-rank particle-hole excitations in a similar but not identical fashion as in other popular many-body methods like coupled-cluster (CC) or in-medium SRG (see e.g.~\cite{Nooijen92} for a connection between GF and CC formalisms).
Importantly, by preserving at each order the analytical properties of the exact self-energy, the ADC expansion ensure that causality is not violated.
Moreover, this form allows the derivation of an energy-independent auxiliary eigenvalue problem that significantly simplifies the numerical solution of Dyson and Gorkov equations, as discussed in Sec.~\ref{sec_working}.
\begin{figure}[b]
\centering
\includegraphics[width=12cm]{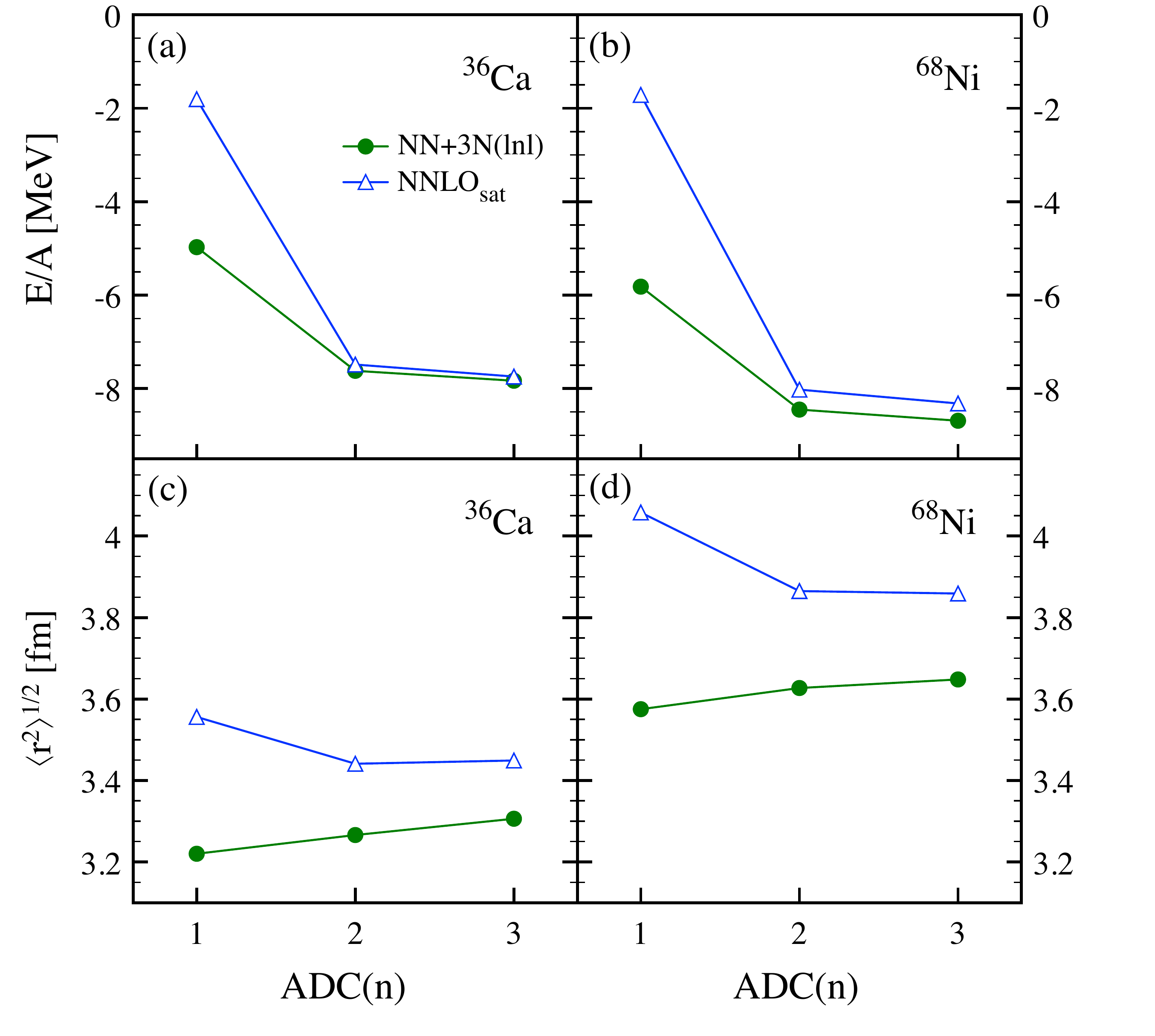}
\caption{
Ground-state energies (top panels) and rms charge radii (bottom panels) of $^{36}$Ca and $^{68}$Ni computed within different ADC($n$) truncation schemes.
Results for the \sat{} and \lnl{} interactions are displayed.
Reprinted figure with permission from Som\`a et al.~\cite{Soma20}, copyright (2020) by the American Physical Society.
}
\label{fig_ADC_convergence}
\end{figure}

The ADC scheme has been developed in the context of finite nuclei in the past few years.
At present, ADC(1), ADC(2) and ADC(3) self-energies are implemented in the Dyson framework~\cite{Cipollone13, Duguet17}, while ADC(1) and ADC(2) are available in the Gorkov case~\cite{Soma11, Soma14a}. 
Gorkov-ADC(3) is under construction within a project whose goal is to automatise the generation of the associated self-energy diagrams.
In Dyson theory, the ADC formalism has been recently generalised to the presence of three-body interactions~\cite{Raimondi18}.

Generally speaking, ADC($n$) defines a truncation scheme that is systematically improvable in the sense that going to higher orders should provide results closer and closer to the exact solution, recovered in the limit of ADC($\infty$).
Nevertheless, in the case of perturbation theory (and, by consequence, resummation methods building on MBPT), there does not exist a well-defined expansion parameter informing on the accuracy associated to a certain truncation level.
Hence, there are two ways of assessing the accuracy of a given approximation: i) by comparing successive orders in the expansion and ii) via benchmarks with exact methods.
Concerning the first possibility, a typical convergence behaviour for ADC(1-3) is shown in Fig.~\ref{fig_ADC_convergence} for ground-state energies and root-mean-square (rms) charge radii of two representative medium-mass nuclei. 
Two different interactions are used, the ``soft'' \lnl{} and the ``harder'' \sat, see Sec.~\ref{sec_interaction} for details.
One observes a clear pattern going from ADC(1) to ADC(3) for all quantities. 
For total energies, while ADC(2) already yields a qualitatively good account, additional correlations introduced at the ADC(3) level are deemed necessary for an accurate description.
In contrast, charge radii are already largely converged at the ADC(2) level.
After analysing the ADC convergence behaviour, one expects ADC(4) corrections to be small for total energies and negligible for charge radii.
The accuracy of the ADC(3) scheme is also confirmed by a direct benchmark against no-core shell model in $^{16}$O, with total energies computed in the two approaches differing by less than $1\%$ for any tested interaction~\cite{Soma20}.

\vskip0.2cm
\subsection{Working equations}
\label{sec_working}
\vskip0.2cm

In practice, solving Dyson or Gorkov equations in their forms \eqref{eq_Dyson} and \eqref{eq_Gorkov} is problematic, specially for finite systems where a solution has to be achieved for numerous (discrete) values of the energy $\omega$.
A method to overcome the problem was proposed once again in quantum chemistry~\cite{Schirmer89}, and consists of two steps.
In the first one, after exploiting the analytical energy dependence of the propagator, Eq.~ \eqref{eq_Dyson} is transformed into an eigenvalue equation where the (energy-dependent) self-energy acts as an effective one-body potential.
The second steps makes use of the analytical energy dependence of the self-energy in the form~\eqref{eq_sigma_Lehmann} and rewrites the problem as an energy-independent eigenvalue equation.
The latter constitutes the working equation to be solved iteratively, and whose solutions directly provide amplitudes and energies entering Eq.~\eqref{eq_Lehmann} for the one-body propagator (see Ref.~\cite{Barbieri17} for a more detailed discussion).
The derivation of this energy-independent eigenvalue problem has been generalised to Gorkov theory in Ref.~\cite{Soma11}.

One disadvantage of the energy-independent formulation is that the number of energy poles, i.e. the eigenvalues and the dimensionality of the energy-independent Dyson or Gorkov matrix, increases at each self-consistent iteration.
In practice, this growth is reduced via the application of Krylov projection techniques~\cite{Schirmer89, Dewul97}, typically implemented by means of a Lanczos algorithm (see Ref.~\cite{Soma14a} for a detailed discussion and a numerical study).

\vskip0.2cm
\subsection{Choice of basis}
\label{sec_basis}
\vskip0.2cm

Equations presented in Sec.~\ref{sec_basics} are general, i.e. are valid in any basis of choice\footnote{Since the present formalism is written in single-nucleon coordinates, the appearance of centre-of-mass (c.o.m.) contributions might pollute the computation of different quantities/observables and must be taken care of. As for energies, the c.o.m. kinetic energy $\hat{T}_\text{c.o.m.}$ is subtracted from the total Hamiltonian from the outset, such that one eventually works with the intrinsic Hamiltonian $\hat{H}_\text{int} \equiv \hat{H} -  \hat{T}_\text{c.o.m.}$. 
Although in the present truncation scheme this does not lead to an exact factorisation of c.o.m. and intrinsic wave functions, it has been demonstrated in similar correlation-expansion methods~\cite{Hagen09, Hergert16} that such an approximate c.o.m. correction is sufficient for all practical purposes.
Furthermore, specific \textit{a posteriori} corrections are applied for radii and densities, see discussion in Sec.~\ref{sec_gs} and, e.g., Ref.~\cite{Cipollone15} for more details.}. 
In an actual calculation, one needs to specify a basis in which operators, together with all relevant quantities, are expanded on.
In the case of atomic nuclei one typically employs a one-body spherical harmonic oscillator (HO) basis, whose eigenfunctions are well suited to the description of a confined system\footnote{On the other hand, the asymptotic behaviour of HO wave functions do not correctly account for the fall-off of the nuclear wave function. As a consequence, a HO basis is not well suited to describe states near the particle continuum, where the long-range part of the wave function is particularly important. In this case, a possibility consists in complementing the HO with a basis specifically designed to account for resonances and non-resonant continuum, e.g. the Berggren basis~\cite{Berggren68, Berggren71}.}.
A HO basis is characterised by two parameters: the oscillator inverse length $\hbar \Omega$ and the number of considered HO wave functions. 
The latter is usually determined by the parameter $e_\text{max} \equiv \text{max}(2n + l)$, which sets the energy threshold of a basis eigenfunction.
Many-body bases are subsequently built as direct products of one-body bases.
While naturally for a $k$-body operator one would set $e_{k\text{max}} = k \cdot e_\text{max}$, the storage of three-body matrix elements for realistic values of $e_\text{max}$ presently constitutes an issue and obliges one to work with $e_{3\text{max}} \ll 3 \cdot e_\text{max}$.
In current state-of-the-art implementations, typical values of $e_\text{max} = 12-15$, $e_{2\text{max}} = 2 \cdot e_\text{max}$ and $e_{3\text{max}}=14-18$ are used.
\begin{figure}
\centering
\includegraphics[width=11cm]{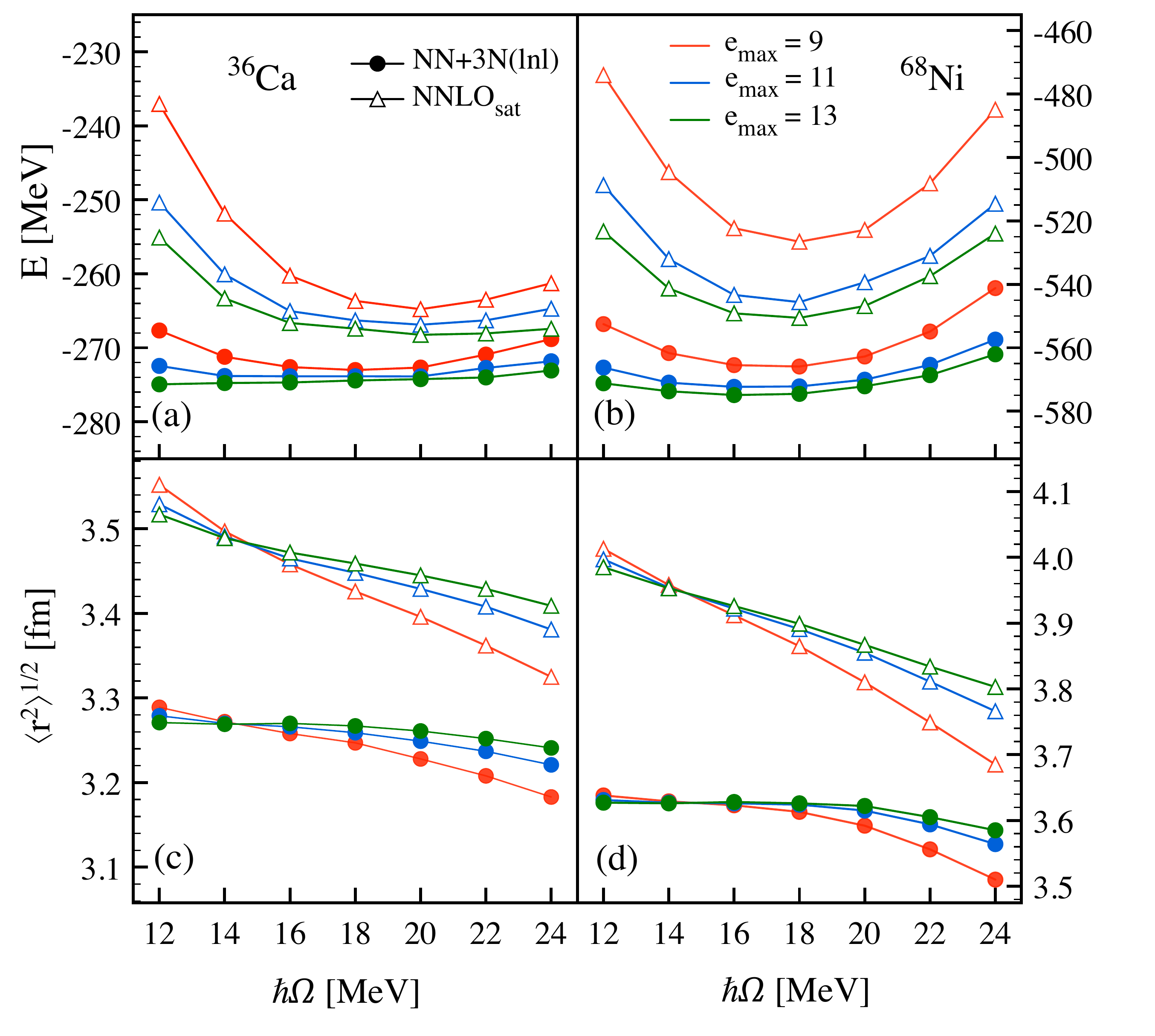}
\caption{
Ground-state energies (top panels) and rms charge radii (bottom panels) of $^{36}$Ca and $^{68}$Ni computed with the \lnl{} and \sat{} Hamiltonians as a function of the harmonic oscillator parameter $\hbar\Omega$ and for increasing size $e_{\text{{\rm max}}}$ of the one-body basis.
In all cases $e_{3\text{max}}=16$ was used.
Reprinted figure with permission from Som\`a et al.~\cite{Soma20}, copyright (2020) by the American Physical Society.
}
\label{fig_Nmax_convergence}
\end{figure}

The first step of a calculation consists in studying the convergence properties (of the observables of interest) with respect to the basis size and oscillator parameter.
An example is shown in Fig.~\ref{fig_Nmax_convergence}, where the basis dependence of ground-state energy and charge radius is investigated for two different nuclei and two different Hamiltonians with ADC(2) calculations.
One sees that in $^{36}$Ca an $e_\text{max} = 13$ model space guarantees sufficiently converged results.
In $^{68}$Ni and for the higher-cutoff Hamiltonian \sat{} (see next section for details) the convergence is not yet optimal.

Combining uncertainties from basis truncation, discussed here, and from many-body expansion, discussed in Sec.~\ref{sec_ADC}, one can evaluate the total theoretical error of the method for a given input interaction.
To give an example~\cite{Soma20}, for soft potentials, like the \ema{} and the \lnl{} Hamiltonians presented in the next section, one estimates an error of about $2\%$ ($<0.5\%$) associated to ground-state energies (radii) of medium-mass nuclei.
For harder potentials, like the \sat{} Hamiltonian, the errors rise to $4\%$ and $1\%$ for energies and radii respectively.
To a good extent, these uncertainties cancel out when differential observables are considered, as in the case of one- or two-nucleon separation energies.
In the results presented in the following such systematic uncertainties are not explicitly reported in the figures (except for Figs.~\ref{fig_lepton} and~\ref{fig_xe}) and should be kept in mind by the reader.

\vskip0.2cm
\subsection{Choice of input interaction}
\label{sec_interaction}
\vskip0.2cm

Like most of ab initio techniques, self-consistent GFs can take in principle any nucleon-nucleon plus three-nucleon (NN+3N) Hamiltonian as input.
In the early 2000s, nearly all applications were performed using semi-phenomenological\footnote{E.g., based on a one-boson exchange model plus phenomenological corrections.} potentials like CD-Bonn~\cite{Machleidt01} or Nijmegen~\cite{Stoks94}, possibly complemented with three-nucleon forces~\cite{Soma08, Soma09}.
Although these interactions had proven successful in the description of light nuclei, their ``hard" character, i.e. the associated large momentum cutoff, required the use of sophisticated resummation schemes, thus hindering applications in medium-mass nuclei.
Starting from 2010, the use of Hamiltonians derived in the context of chiral effective field theory ($\chi$-EFT)~\cite{Hammer20} began to spread.
Compared to the previous phenomenological models, $\chi$-EFT interactions present several advantages:
\begin{enumerate}
\item By explicitly taking into account only low-energy degrees of freedom, they have a much smaller associated cutoff;
\item All many-body (i.e., 2N, 3N, 4N, ...) operators and currents are derived consistently without any ad-hoc assumption, which augments the predictive power;
\item A theoretical error can be associated to a given $\chi$-EFT Hamiltonian (which relates to the employed truncation in the EFT expansion).
\end{enumerate}

Nowadays, $\chi$-EFT interactions constitute the standard for ab initio nuclear structure calculations.
The first feature, i.e. their ``softness'', is often further amplified by the use of similarity renormalisation group (SRG) techniques~\cite{Bogner10}, i.e. unitary transformations of the Hamiltonian that further decouple low- and high-momentum modes, leading to much improved convergence properties.
The third point implies that the EFT error can (and should) be subsequently propagated to many-body observables. 
This has been done in practice only very recently (see e.g. Refs.~\cite{Epelbaum15a, Binder18, Drischler19, Huther19, Epelbaum19a, Epelbaum19b, Tichai20}) and more formal and technical developments along these lines will be required in the future.

Mainly three different $\chi$-EFT Hamiltonians have been employed in recent GF calculations, all of which are discussed here.
The first one, labelled \ema, is based on the next-to-next-to-next-to-leading order (N$^3$LO) nucleon-nucleon potential from Entem and Machleidt~\cite{Entem03,Machleidt11} combined with the N$^2$LO $3N$ interaction with a local regulator~\cite{Navratil07}. The $2N$ interaction of Ref.~\cite{Entem03} was built with a cutoff of 500 MeV/c, however, a 400 MeV/c regulator was used for the $3N$ sector~\cite{Roth12}.
This Hamiltonian has been systematically applied to $p$- and $sd$-shell nuclei and yields a good reproduction of oxygen, nitrogen and fluorine binding enrgies~\cite{Hergert13, Cipollone13, Cipollone15}. 
Nevertheless, it leads to overbinding in medium-mass nuclei starting in the calcium chain and underpredicts nuclear radii even for O isotopes~\cite{Binder14, Soma14b, Lapoux16}. 

With the main objective of improving on the description of radii, a chiral Hamiltonian with terms up to N$^2$LO was developed in Ref.~\cite{Ekstrom15}.
It is characterised by a simultaneous fit of $2N$ and $3N$ LECs that does not rely solely on two-nucleon and $A=3,4$ data, but also on binding energies of $^{14}$C and $^{16,22,24,25}$O as well as charge radii of $^{14}$C and $^{16}$O. 
The resulting interaction, named \sat, successfully describes the saturation of infinite nuclear matter~\cite{Ekstrom15} as well as various observables in mid-mass nuclei, including charge radii~\cite{Lapoux16, Raimondi19a, Idini19, Barbieri19}.
Unlike the \ema{} interaction, \sat{} employs a non-local regulator.

Motivated by the success of \sat, a novel interaction named \lnl{} was presented recently~\cite{Soma20}.
The goal was to amend the original \ema{} interaction, and in particular its $3N$ part.
While the latter has been shown to be problematic, its $2N$ part is instead believed to perform relatively well and thus is kept unchanged.
Being based on the N$^3$LO potential, which provides a better description of nucleon-nucleon data compared to the lower-order \sat, it yields superior features in light systems, e.g. a better reproduction of spectroscopy of natural parity states in $p$- and light $sd$-shell nuclei. 
\lnl{} has been also shown to provide a very good description of energy observables (ground-state energies and energy spectra) in medium-mass nuclei up to mass $A \sim 60$~\cite{Soma20}.

\vskip0.2cm
\subsection{Computational requirements}
\label{sec_numerical}
\vskip0.2cm

As any other state-of-the-art ab initio nuclear structure approach, the SCGF method requires the development of an advanced numerical code.
The computational cost of a simulation strongly depends on (i) the size of the model space, (iii) the chosen level of approximation\footnote{For a given approximation, a Gorkov calculation is more costly than a Dyson one because of the increased dimensionality of the quasiparticle space. Estimates presented in this section refer to Dyson calculations of typical medium-mass nuclei.} and (iii) the two- or three-body character of the input Hamiltonian.
ADC(1) calculations with only NN forces can be easily performed on a laptop also in large bases (in few CPU minutes).
Going to ADC(2) limits a laptop calculation to a small, yet (semi-)realistic model space, typically $e_\text{max} = 8-9$ (doable in a few CPU hours).
If larger bases are needed (to ensure model-space convergence, e.g. typically $e_\text{max} = 12-13$), then one has to resort to a dedicated computer cluster (with a corresponding cost of few hundred CPU hours).
ADC(3) is doable on a laptop only for very small model spaces and any realistic calculation requires the implementation of MPI parallelisation and the use of a high-performance computing centre (with running times of several thousand CPU hours). 

The inclusion of 3N forces results into an increase of both CPU time (due to the higher rank of the tensors at play) and memory usage (due to the larger amount of matrix elements to be stored).
As a consequence, on a laptop and for realistic bases, even an ADC(1) calculation becomes heavy in terms of CPU and one quickly reaches the limits in terms of available RAM\footnote{See also discussion in Sec.~\ref{sec_heavy} on storage of 3N matrix elements.}.
ADC(2) and ADC(3) calculations require optimised implementations and the use of a high-performance computing centre, with typical running times of a few thousand and tens of thousands CPU hours respectively.

\vskip0.2cm
\section{Recent applications}
\label{sec_applications}
\vskip0.2cm

\vskip0.2cm
\subsection{Ground-state properties}
\label{sec_gs}
\vskip0.2cm

\begin{figure}
\centering
\includegraphics[width=8.2cm]{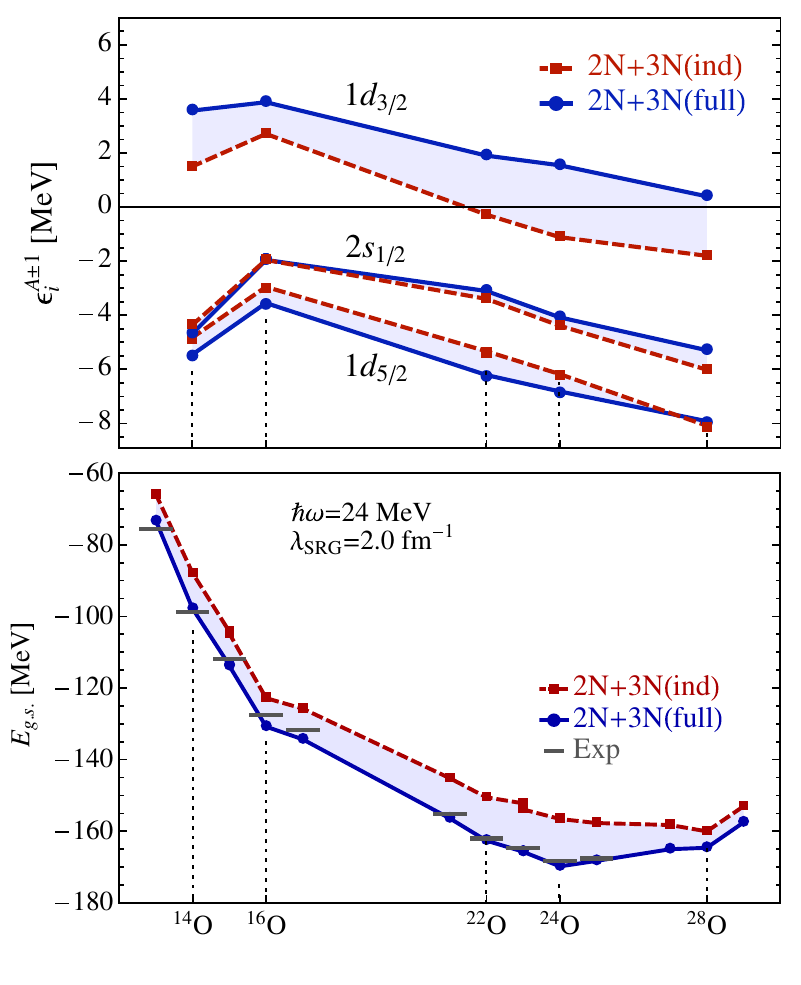}
\hfill
\includegraphics[width=9cm]{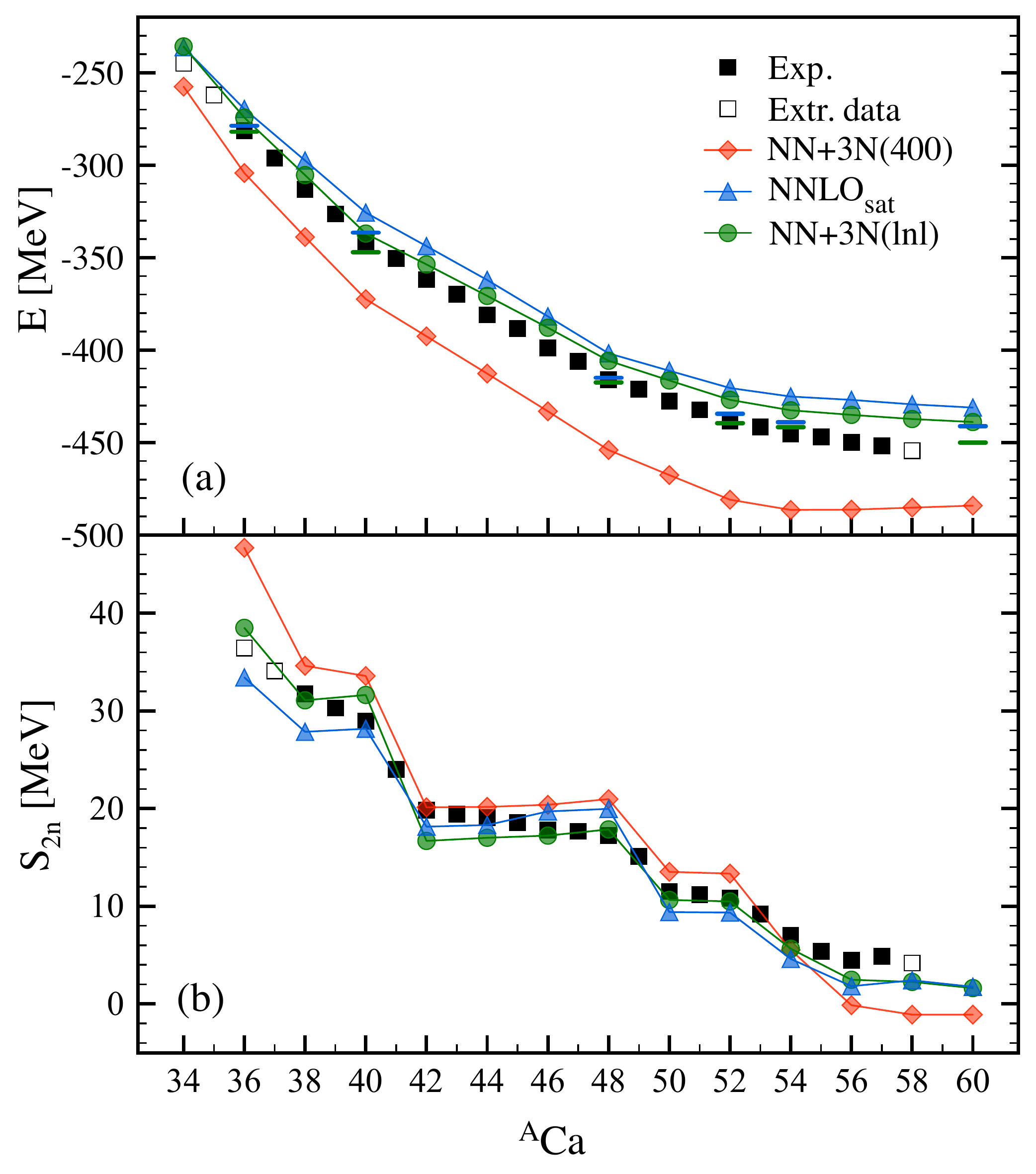}
\caption{
\textit{Left:} 
ADC(3) calculations of closed-shell oxygen isotopes performed with the \ema{} Hamiltonian (blue points and lines).
For reference, results with only original $2N$ operators are displayed (red points and lines).
\textit{Top:}
One-neutron addition and removal energies associated to dominant quasiparticle peaks.
\textit{Bottom:}
Ground-state energies compared with experimental values (grey bars). 
Adapted figure with permission from Cipollone et al.~\cite{Cipollone13}, copyright (2013) by the American Physical Society.
\textit{Right:} 
ADC(2) calculations of even-even calcium isotopes performed with the \ema, \sat{} and \lnl{} Hamiltonians (coloured points and lines), compared with measured and extrapolated data (black points).
ADC(3) results are depicted as horizontal lines when available.
\textit{Top:}
Total ground-state energies.
\textit{Bottom:}
Two-neutron separation energies.
Adapted figure with permission from Som\`a et al.~\cite{Soma20}, copyright (2020) by the American Physical Society.
}
\label{fig_BE}
\end{figure}

The total ground-state energy, or binding energy, of a nucleus constitutes the most basic nuclear structure observable.
In Green's function theory, total energies are preferably computed via the generalised GMK sum-rule~\eqref{eq_Koltun_W}.
While earlier applications made use of a $2N$-only Hamiltonian, possibly complemented by a phenomenological correction to compensate for missing $3N$~\cite{Barbieri09a,Barbieri09b,Soma13}, starting from 2013 calculations with realistic $2N+3N$ interactions could be routinely performed.
A representative example concerns the oxygen chain~\cite{Cipollone13} and is shown in Fig.~\ref{fig_BE} (left). 
In the bottom panel, ADC(3) ground-state energies are displayed for closed-shell oxygen isotopes, computed with the \ema{} interaction respectively excluding and including original $3N$ operators\footnote{The SRG evolution described in Sec.~\ref{sec_interaction} and standardly applied to nuclear Hamiltonians induces additional many-body operators that need to be taken into account~\cite{Bogner10}. Hence, one has to distinguish between \textit{original} and \textit{induced} e.g. $3N$ forces.}.
One notices that the addition of $3N$ forces is crucial for a quantitative reproduction of experimental data.
In particular, when only a $2N$ interaction is considered, the neutron dripline is wrongly located at $N=20$, while it is correctly reproduced at $^{24}$O in the presence of $3N$ forces.

In such a context it can be instructive to inspect one-neutron addition and removal energies associated to dominant quasiparticle peaks, see top-left panel of Fig.~\ref{fig_BE}.
One sees that the $3/2^+$ fragment becomes bound in neutron-rich isotopes when the $2N$-only Hamiltonian is employed.
When $3N$ interactions are switched on, it is instead pushed up and remains unbound all the way to $^{28}$O, thus explaining the position of the dripline.
This observation confirmed the repulsive character of the Fujita-Miyazawa $3N$ interaction, as previously discussed in Ref.~\cite{Otsuka10}.

This result was one of the first successful applications of ab initio techniques beyond light nuclei.
The oxygen chain also constituted, for a few years, a testbed where calculations from various approaches could be benchmarked, demonstrating the reliability of the different many-body truncations~\cite{Hergert13,Epelbaum14,Jansen14,Hebeler15}.
More recently, the availability of new Hamiltonians prompted calculations of heavier systems, from calcium up to the nickel chain.
An example is constituted by Gorkov GF calculations exploring the performance of three interactions along few semi-magic chains, as reported in Ref.~\cite{Soma20}.
A typical outcome is the one displayed in Fig.~\ref{fig_BE} (right), where ground-state energies of calcium isotopes as well as their differences, two-neutron separation energies, are shown as a function of neutron number.
The overbinding generated by the older \ema{} clearly stands out.
In contrast, the newer Hamiltonians \sat{} and \lnl{} yield an excellent reproduction of total as well as differential ground-state energies, specially once ADC(3) corrections are taken into account.
A good performance is found also in the nickel chain, up to the point where current computational limitations hinder a complete model-space convergence of the calculations~\cite{Soma20}.
In addition to semi-magic chains, the theory was tested in relation to novel experimental measurements in potassium~\cite{Rosenbusch15}, titanium~\cite{Leistenschneider18} and argon~\cite{Mougeot20} chains.

The performance of these three Hamiltonians was further investigated on nuclear radii in Ref.~\cite{Soma20}.
It was found that \sat{} provides a good account of rms charge radii all the way up to nickel.
Specifically, the bulk contributions are well described already at the ADC(2) level, while finer details (e.g. the parabolic behaviour observed between $^{40}$Ca and $^{48}$Ca) need further improvement of the many-body truncation and/or the interaction.
Density distributions provide even further insight into the way nucleons arrange themselves in the correlated nuclear medium.
The nuclear charge density distribution is typically obtained as a sum of three contributions~\cite{Brown79},
\begin{equation}\label{charge_density}
\rho_{{\rm ch}}(r) = \rho_{{\rm ch}}^{{\rm p}}(r) + \rho_{{\rm ch}}^{{\rm n}}(r) + \rho_{{\rm ch}}^{{\rm ls}}(r),
\end{equation}
where $\rho_{{\rm ch}}^{{\rm p}}$ ($\rho_{{\rm ch}}^{{\rm n}}$) is determined by folding the point-proton (point-neutron) density with the finite charge distribution of the proton (neutron) and $\rho_{{\rm ch}}^{{\rm ls}}$ is a relativistic correction that depends on spin-orbit terms.
In addition, centre-of-mass and relativistic Darwin-Foldy corrections are taken into account by employing an effective position variable~\cite{negele70}.
Finite nucleon charge distributions can be expressed as a sum of Gaussians, with the parameters adjusted to reproduce form factors from electron scattering data~\cite{chandra76a}.
The relativistic spin-orbit correction is usually computed within the factorisation approximation introduced in Ref.~\cite{Bertozzi72}.
Even though $\rho_{{\rm ch}}^{{\rm p}}$ largely dominates, the other two contributions can visibly alter the total charge distribution in some cases.

\begin{figure}[t]
\centering
\includegraphics[width=.48\textwidth]{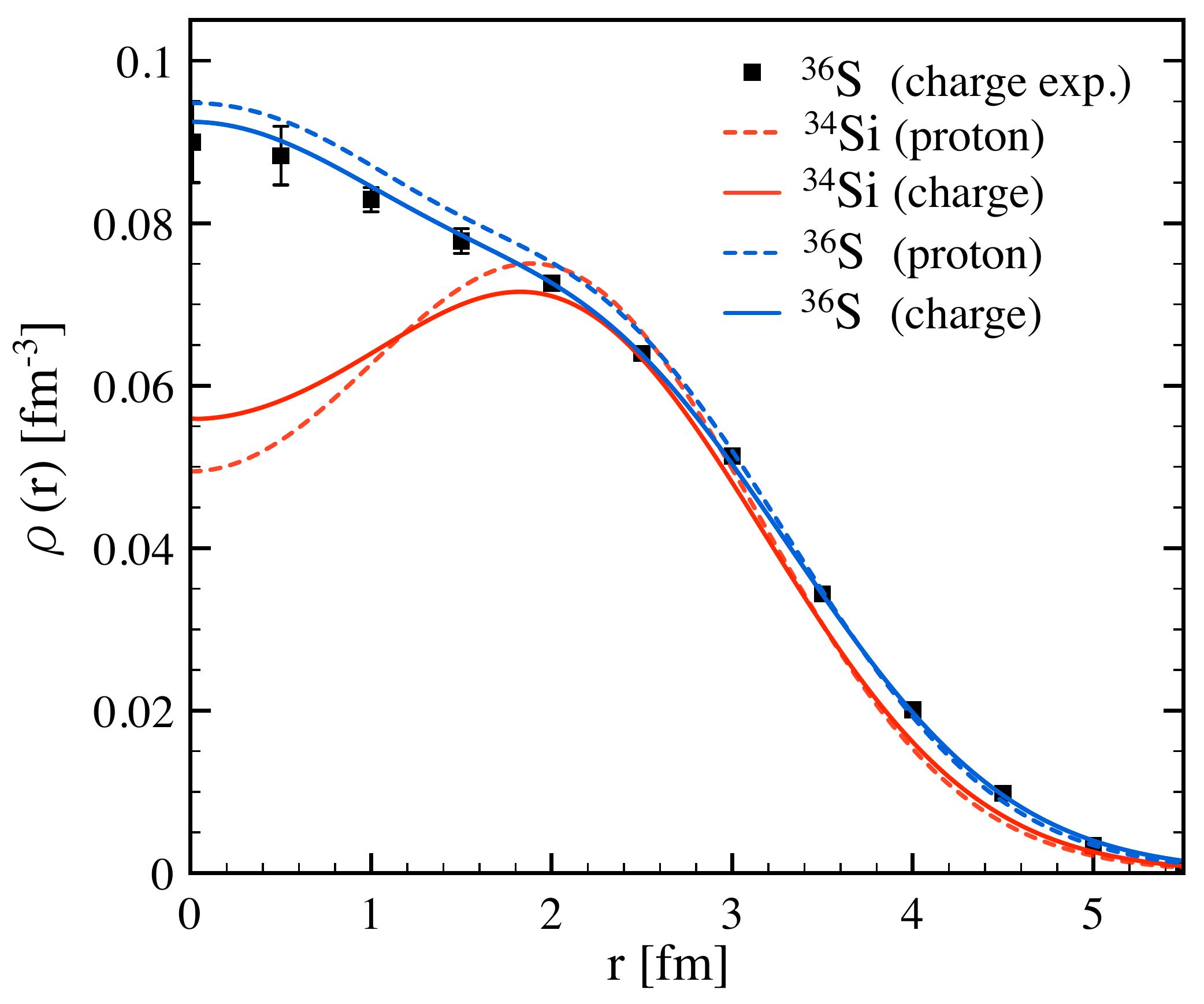}
\hfill
\includegraphics[width=.48\textwidth]{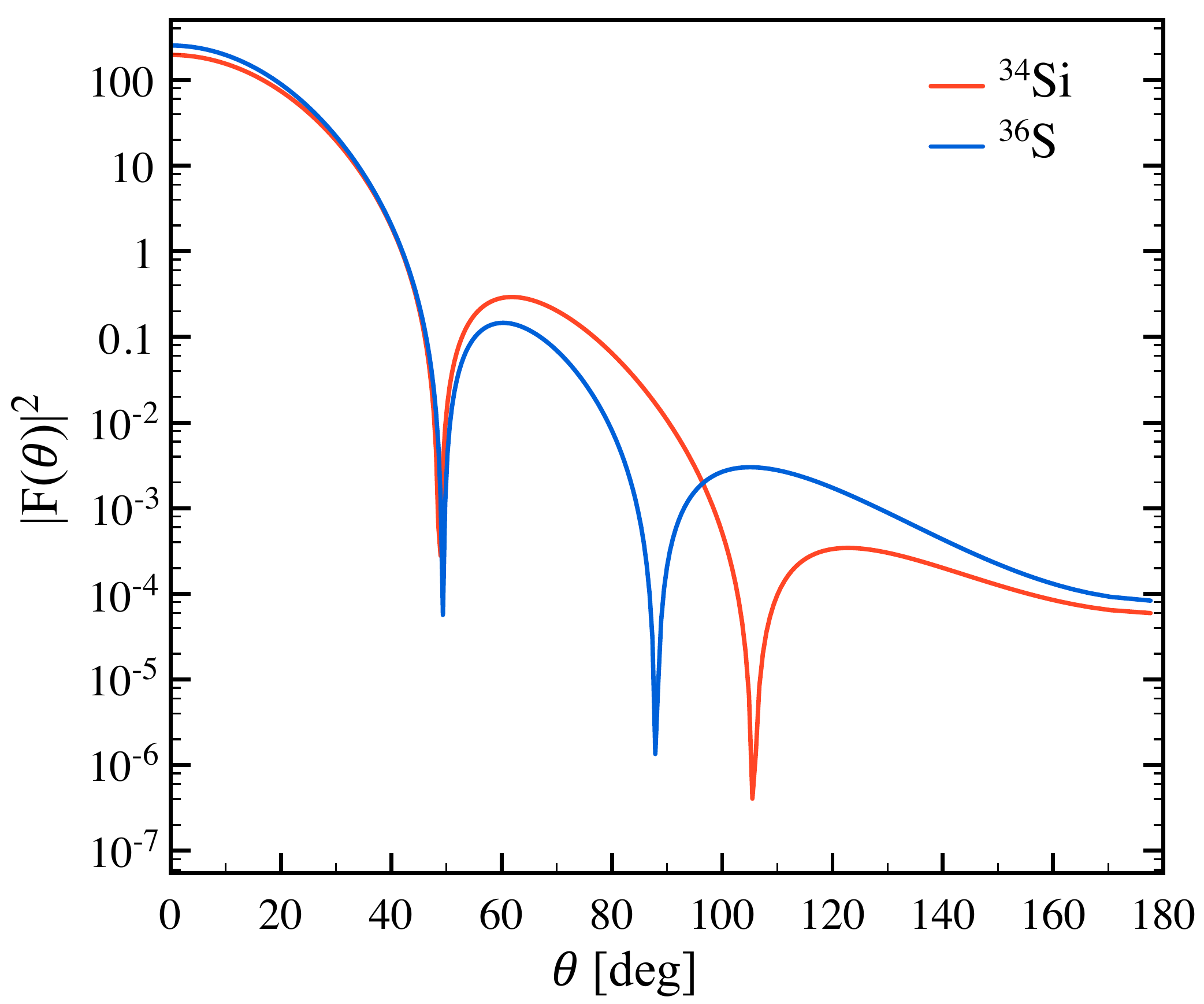}
\caption{
\textit{Left:} 
Point-proton and charge densities of $^{34}$Si and $^{36}$S computed at the ADC(3) level with the \sat{} interaction.
The experimental charge density of $^{36}$S is taken from Ref.~\cite{Rychel83}.
\textit{Right:} 
Angular dependence of the charge form factor computed for 300 MeV electron scattering.
Modified figure with permission from Duguet et al.~\cite{Duguet17}, copyright (2017) by the American Physical Society.
}
\label{fig_densities}
\end{figure}

As for radii, densities are computed directly from the one-body GF and can therefore be routinely evaluated for medium-mass systems.
An interesting example relates to the possible presence of a depletion in the central part of the charge density profile, usually referred to as \textit{bubble}. 
One of the most likely candidates has been identified in the nucleus $^{34}$Si~\cite{richter03a,ToddRutel04,Khan07,Grasso09,Yao12,Yao13,Wu14}.
In Ref.~\cite{Duguet17} this system, together with its $Z+2$ partner $^{36}$S, has been thoroughly investigated by means of both Dyson and Gorkov GF calculations.
The resulting point-proton and charge density distributions are shown in Fig.~\ref{fig_densities} (left).
The effect of folding with the finite size of the proton is specially visible in the centre, with an attenuation of oscillations in the charge profiles. 
The agreement between the computed and the measured charge distribution for the stable nucleus $^{36}$S is excellent, which gives confidence in the prediction for the unstable $^{34}$Si, yet unknown experimentally.
For the latter, a depletion in the region below 2 fm is indeed found. 
Its magnitude is comparable or even larger than what previously obtained in EDF~\cite{Yao12,Yao13} or shell model calculations~\cite{Grasso09}.

A measurement of the charge distribution of $^{34}$Si would require electron scattering on radioactive ions. 
These experiments are becoming feasible only now~\cite{Suda17}, with first results on the heavier $^{132}$Xe obtained by the SCRIT collaboration~\cite{Tsukada17}.
In the coming years, the case of $^{34}$Si will thus constitute an interesting objective for electron-nucleus scattering facilities.
Such a measurement would extract the electromagnetic charge form factor, related to the charge profile via
\begin{equation}
F(q) = \int d\vec{r} \rho_{{\rm ch}}(r)e^{-i\vec{q} \cdot \vec{r}}\, ,
\end{equation} 
where $\vec{q}$ is the transferred momentum, itself related to the incident momentum $\vec{p}$ and the scattering angle $\theta$ via $q = 2p\sin\theta/2$. 
The calculated charge form factors for 300 MeV electron scattering on $^{34}$Si and $^{36}$Si are displayed in Fig.~\ref{fig_densities} (right).
Clear differences appear in the angular dependence for the two systems, with a higher magnitude and a displaced position of the second minimum for $^{34}$Si.
This analysis gives indications on what range of transferred momenta, and consequently which luminosities, are necessary for identifying possible depleted density profiles in this mass region.

\vskip0.2cm
\subsection{Excited-state properties}
\label{sec_excited}
\vskip0.2cm

\begin{figure}
\centering
\includegraphics[width=8cm]{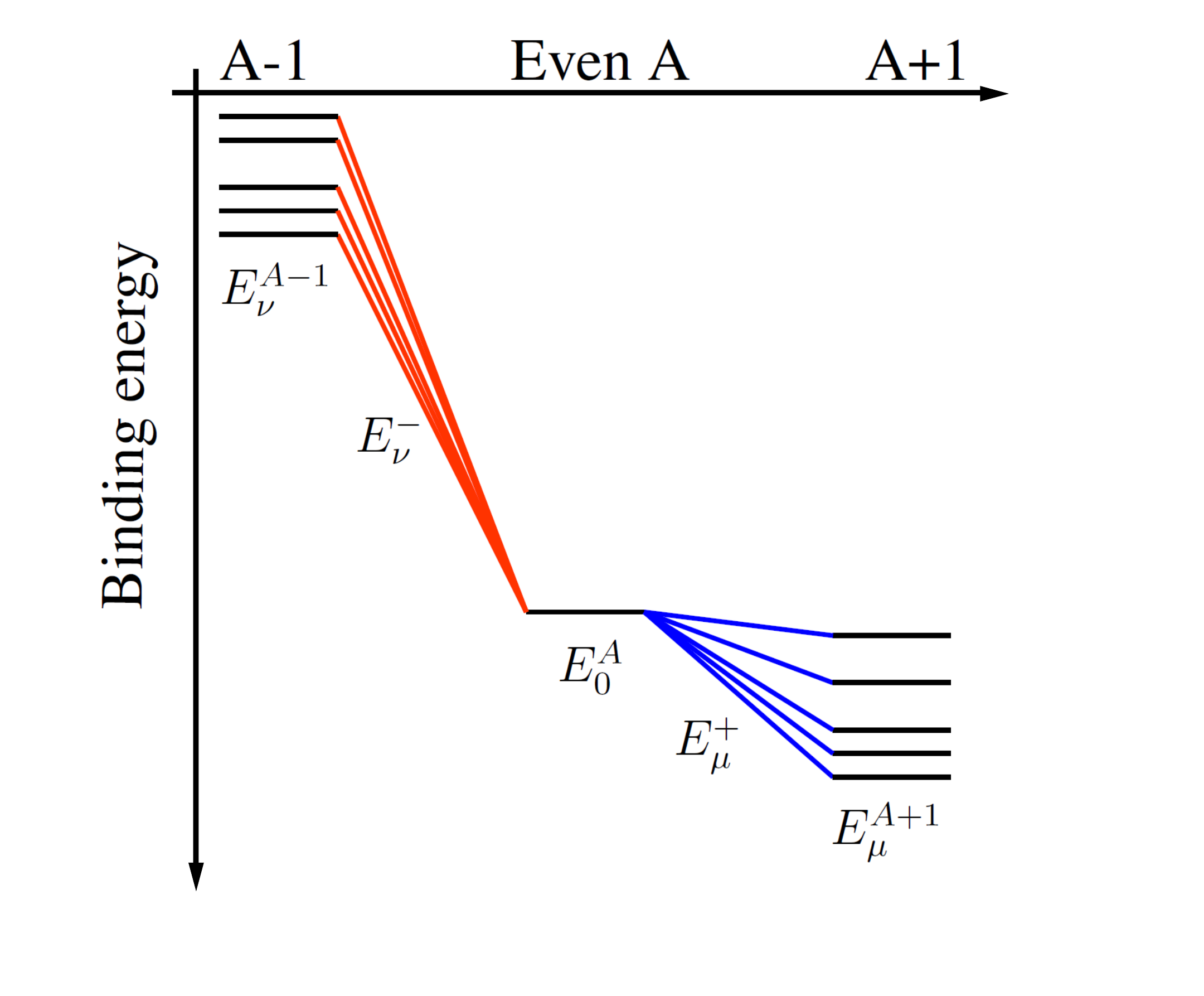}
\quad
\quad
\includegraphics[width=8.5cm]{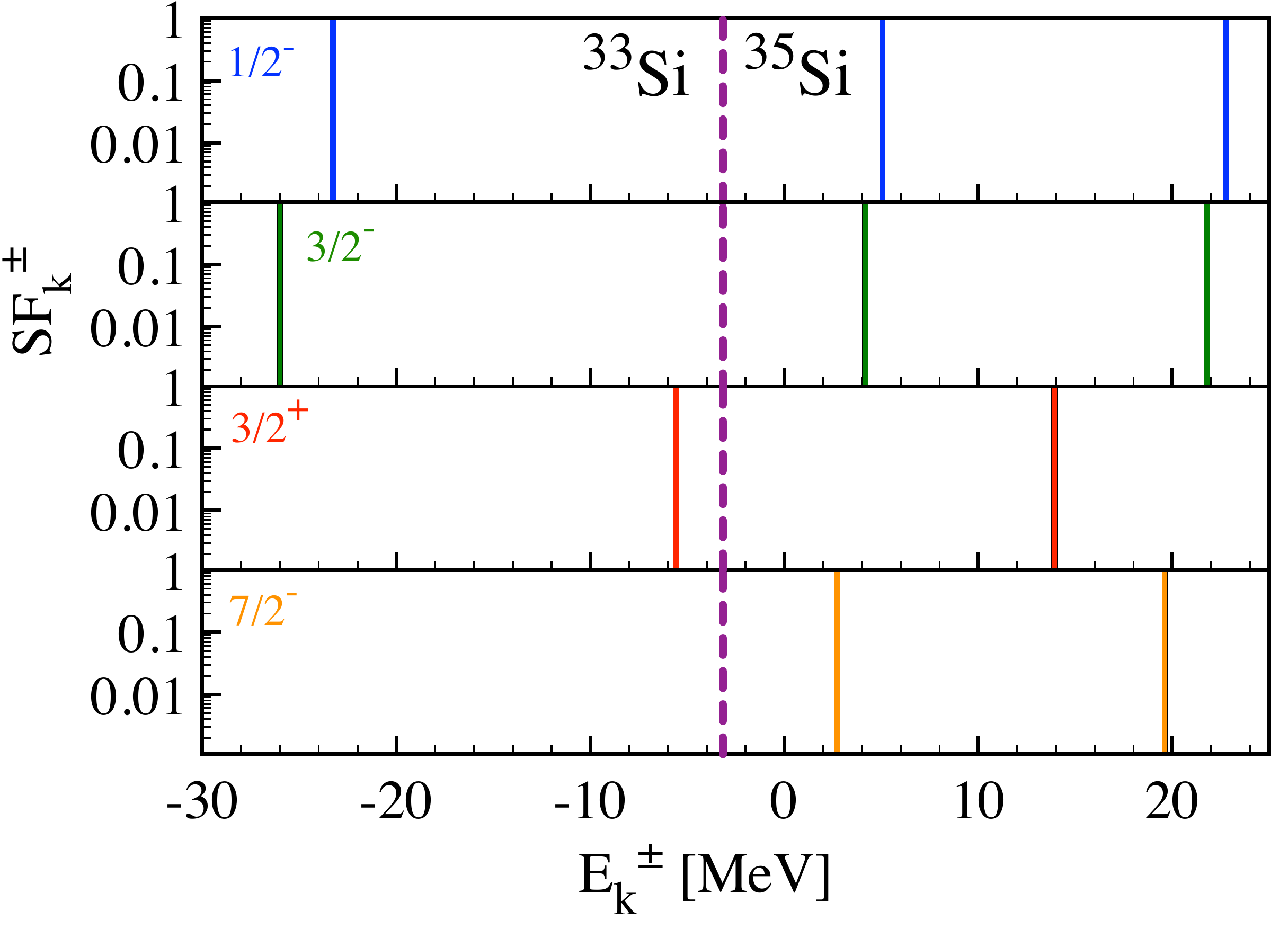}
\includegraphics[width=8.5cm]{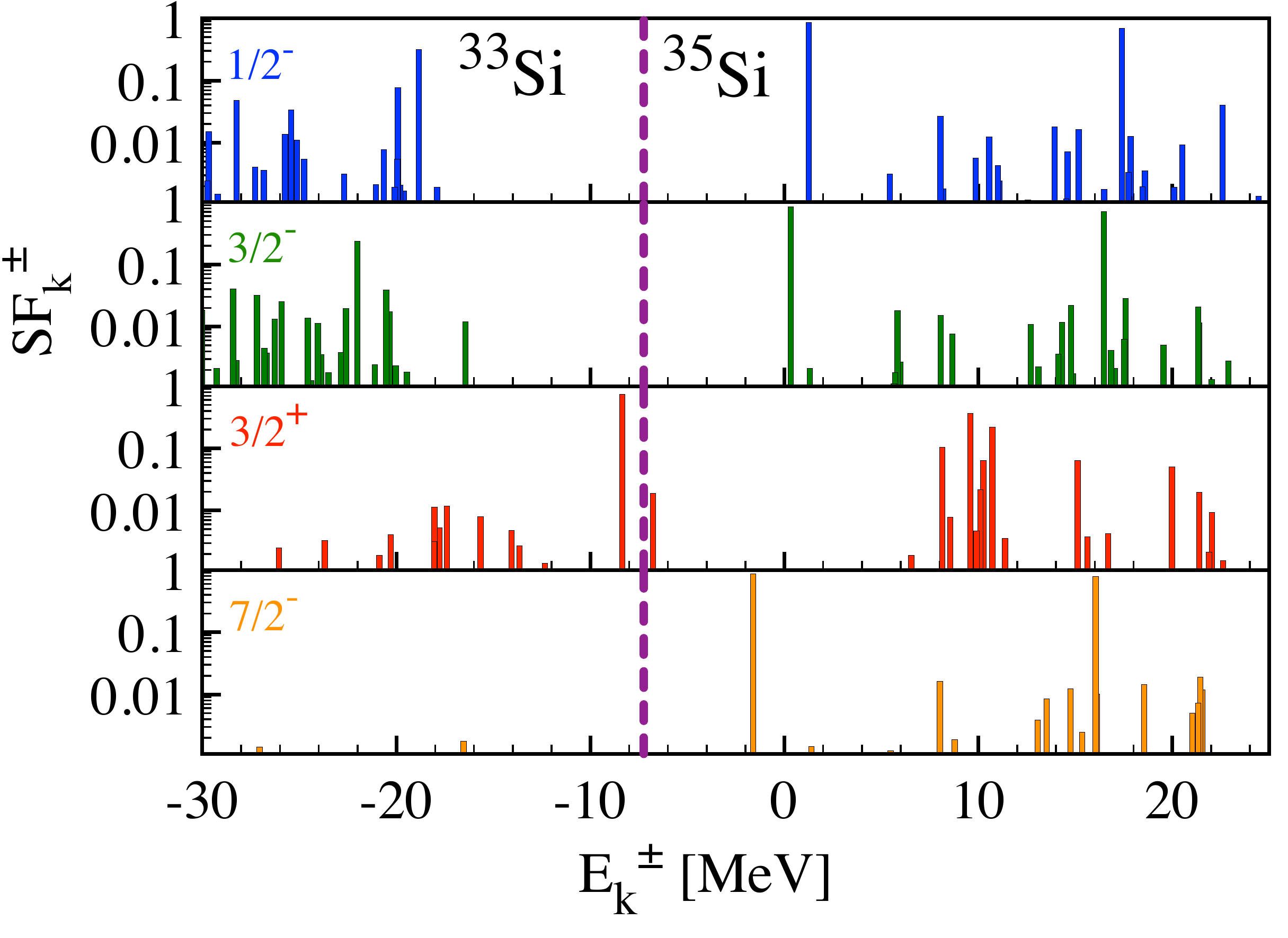}
\quad
\includegraphics[width=8.5cm]{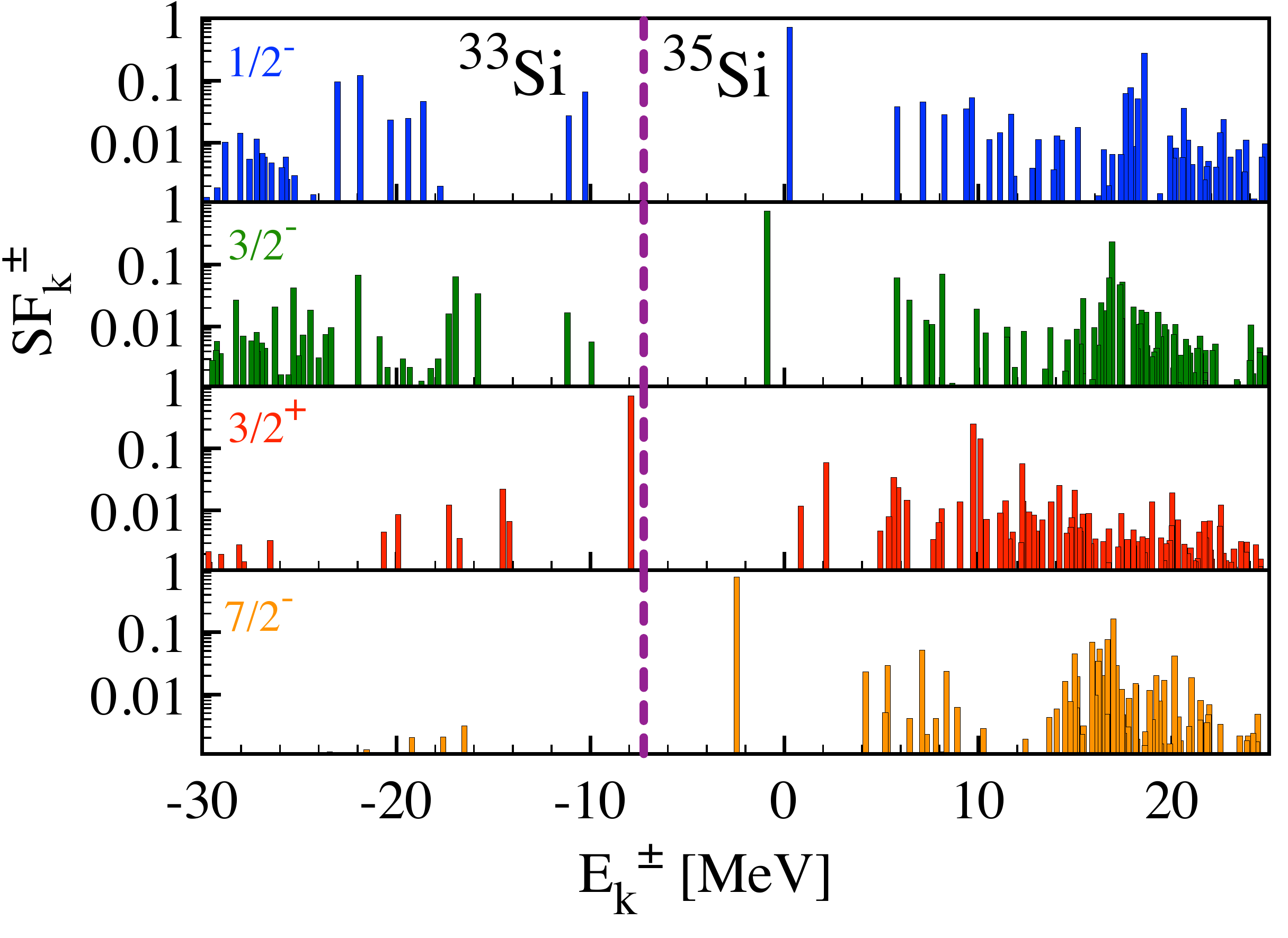}
\caption{
\textit{Top left:} 
Schematic view of the $A \pm 1$ states (and associated energies) reached via one-nucleon addition and removal.
\textit{Top right:}
Neutron spectral strength distribution in $^{34}$Si computed at the ADC(1) level with the \sat{} interaction.
The four panels display different angular momentum and parity channels.
The dashed vertical line represents the Fermi energy, separating the spectra of $^{33}$Si and $^{35}$Si.
\textit{Bottom:} 
Same as above but for ADC(2) and ADC(3) calculations.
Data originates from the results of Ref.~\cite{Duguet17}.
}
\label{fig_spectra_Si}
\end{figure}

An asset of GF theory resides in the rich content of the one-body propagator, which does not solely provide information  on the targeted (even-even) system with mass number $A$ but also on the four neighbouring (odd-even) $A\pm1$ nuclei.
This information is explicit in the Lehmann representation of the GF, Eq.~\eqref{eq_Lehmann}.
The poles $E_i^{\pm}$ of the function~\eqref{eq_Lehmann} correspond to one-nucleon addition and removal energies, as schematically depicted in Fig.~\ref{fig_spectra_Si} (top left).
In addition, the associated amplitudes in the numerator represent the probabilities to reach a specific eigenstate $\ket {\Psi^{\text{A+1}}_n}$ ($\ket {\Psi^{\text{A-1}}_{k}}$) of the $A+1$ ($A-1$) system by adding (removing) a nucleon in a single-particle state to (from) the ground state $\ket{\Psi^{\text{A}}_{0}}$ of the even-even system. 
Those amplitudes can be expanded in a single-particle basis $\{\ac {\alpha}\}$ according to
\begin{subequations}
\label{eq:defu}
\begin{eqnarray}
U^{\alpha}_{n} &\equiv& \bra {\Psi^{\text{A}}_{0}} a_\alpha \ket {\Psi^{\text{A+1}}_{n}}  \, , \\
V^{\alpha}_{k} &\equiv& \bra {\Psi^{\text{A}}_{0}} a^\dagger_\alpha \ket {\Psi^{\text{A-1}}_{k}} \,  .
\end{eqnarray}
\end{subequations}
Next, spectroscopic probability matrices for the nucleon addition and removal can be built\footnote{Here bold symbols denote matrices in the one-body Hilbert space ${\mathcal{H}}_1$.}, $\mathbf{S}_{n}^{+}\equiv \mathbf{U}_{n} \mathbf{U}^{\dagger}_{n}$ and $\mathbf{S}_{k}^{-}\equiv \mathbf{V}^{\ast}_{k}\mathbf{V}^{T}_{k}$, respectively. 
Their elements read as
\begin{subequations}
\label{spectroproba}
\begin{eqnarray}
S_{n}^{+\alpha\beta} &\equiv&  \bra {\Psi^{\text{A}}_{0}} a_\alpha \ket {\Psi^{\text{A+1}}_{n}} \bra {\Psi^{\text{A+1}}_{n}} a^\dagger_\beta \ket {\Psi^{\text{A}}_{0}}    \, \, \, , \label{spectroprobaplus} \\
S_{k}^{-\alpha\beta} &\equiv& \bra {\Psi^{\text{A}}_{0}} a^\dagger_\beta \ket {\Psi^{\text{A-1}}_{k}} \bra {\Psi^{\text{A-1}}_{k}} a_\alpha \ket {\Psi^{\text{A}}_{0}}    \, \, \, . \label{spectroprobamoins}
\end{eqnarray}
\end{subequations}
Taking the trace over the one-body Hilbert space ${\mathcal{H}}_1$ leads to spectroscopic factors
\begin{subequations}
\label{spectrofactor}
\begin{eqnarray}
SF_{n}^{+} &\equiv& \text{Tr}_{{\mathcal{H}}_{1}}\!\left[ \mathbf{S}_{n}^{+}\right] =  \sum_{\alpha \in {\mathcal{H}}_{1}} \left|U^{\alpha}_{n}\right|^2 \, \, , \\
SF_{k}^{-} &\equiv& \text{Tr}_{{\mathcal{H}}_{1}}\!\left[\mathbf{S}_{k}^{-} \right] = \sum_ {\alpha \in {\mathcal{H}}_{1}} \left|V^{\alpha}_{k}\right|^2  \,\, ,
\end{eqnarray}
\end{subequations}
which are the norms of the spectroscopic amplitudes. 
A spectroscopic factor thus sums the probabilities that an eigenstate of the $A+1$ ($A-1$) system can be described as a nucleon added to (removed from) a single-particle state on top of the ground state of the $A$-nucleon system.

The complete spectroscopic information associated with one-nucleon addition and removal processes can be collected into the spectral function $\mathbf{S}(\omega)$, defined as the energy-\emph{dependent} matrix on ${\mathcal{H}}_1$
\begin{eqnarray}
\mathbf{S}(\omega) &\equiv& \!\!\!\!\! \sum_{n \in {\mathcal{H}}_{A\!+\!1}} \!\!\! \mathbf{S}_{n}^{+} \,\, \delta(\omega -E_{n}^{+}) +  \!\!\!\!\!\sum_{k \in {\mathcal{H}}_{A\!-\!1}} \!\!\! \mathbf{S}_{k}^{-}  \,\, \delta(\omega -E_{k}^{-}) , 
\label{eq_spectral}
\end{eqnarray}
where the first (second) sum runs over eigenstates of $H$ in the Hilbert space ${\mathcal{H}}_{A\!+\!1}$ (${\mathcal{H}}_{A\!-\!1}$) associated with the $A+1$ ($A-1$) system. 
Taking the trace of $\mathbf{S}(\omega)$ gives the spectral strength distribution (SDD)
\begin{eqnarray}
{\mathcal{S}}(\omega) &\equiv& \text{Tr}_{{\mathcal{H}}_{1}}\!\left[\mathbf{S}(\omega)\right]  \label{SDD_def} \nonumber \\ 
&=& \! \!\!\!\!   \sum_{n \in {\mathcal{H}}_{A\!+\!1}} \!\!\!\!  SF_{n}^{+} \, \delta(\omega -E_{n}^{+}) + \sum_{k \in {\mathcal{H}}_{A\!-\!1}} \!\!\!\!    SF_{k}^{-}  \, \delta(\omega -E_{k}^{-}) \, ,
\label{eq_sdd}
\end{eqnarray}
which is a basis-independent function of the energy. 
Equations \eqref{eq:defu}-\eqref{eq_sdd} can be generalised to the Gorkov formalism~\cite{Soma11}. 

An example of SDD computed at three different levels of approximation, ADC(1-3), is shown in Fig.~\ref{fig_spectra_Si} for the nucleus $^{34}$Si.
ADC(1), i.e. Hartree-Fock, is a mean-field (or independent-particle) approximation, which translates into a series of quasiparticle peaks with unity spectroscopic factors, i.e. there is a one-to-one correspondence between the single-particle basis $\{\ac {\alpha}\}$ and the many-body states $\ket {\Psi^{\text{A$\pm$1}}_i}$ accessible via the process of adding or removing a nucleon.
One can nevertheless identify the main qualitative features of the SDD, e.g. associate the ground state of $^{33}$Si ($^{35}$Si) with the first peak on the left (right) of the Fermi energy characterised by spin and parity $3/2^+$ ($7/2^-$).
ADC(2) introduces the lowest-order dynamical correlations that lead to a first fragmentation of the spectral distribution.
A number of fragments with small spectroscopic factors appear in the vicinity of the ADC(1) peak, which is now shifted in energy and reduced in strength, i.e. it has $SF_{i}^{\pm} < 1$.
ADC(3) correlations further fragment the quasiparticle strength, giving rise to a large number of small peaks and a further reduction of the main-peak spectroscopic factor.
One notices that, around the Fermi energy, fragments with a good quasiparticle character, i.e. large spectroscopic factors, survive, in accordance to Landau's Fermi liquid theory.
In contrast, away from the the Fermi level the strength is spread over a wide energy interval and one can hardly identify single-particle-like excitations.
\begin{figure}
\centering
\includegraphics[width=18cm]{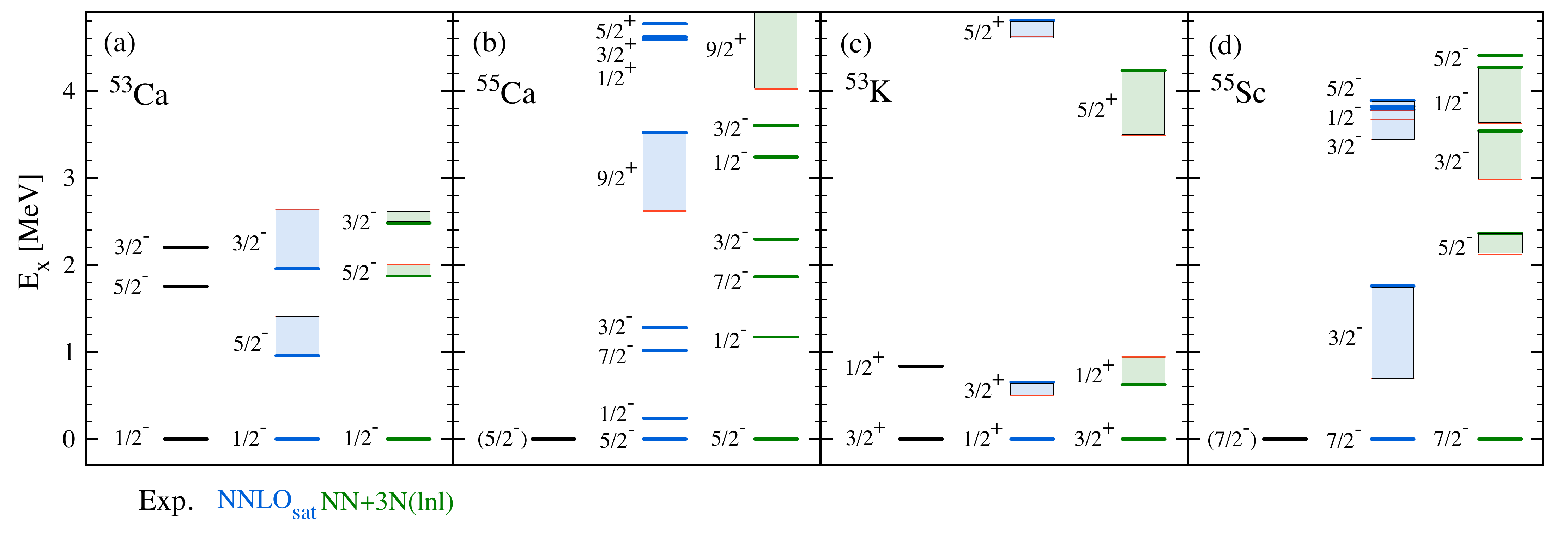}
\caption{
One-nucleon addition and removal spectra from $^{54}$Ca.
Available experimental data appear in the left column of each panel.
GF calculations performed with the \sat{} and \lnl{} interactions are displayed in the central and right column of each panel respectively.
Red lines constitute results obtained at the ADC(3) level. 
Shaded areas connect ADC(2) and ADC(3) where available.
In all panels, states with $E_x < 5$ MeV and $SF < 10 \%$ are shown.
Modified figure with permission from Som\`a et al.~\cite{Soma20}, copyright (2020) by the American Physical Society.
Additional new experimental data are from Refs.~\cite{Chen19, Sun20}.
}
\label{fig_spectra_CaK}
\end{figure}

From the SDD depicted in Fig.~\ref{fig_spectra_Si} one can extract an excitation spectrum of the $A-1$ ($A+1$) system, by looking at increasingly negative (increasingly positive) energies on the left (right) of the Fermi surface (see top-left panel for a schematic representation).
By comparing ADC(2) and ADC(3) truncations, one concludes that the latter is necessary for an accurate description of the main energy peaks~\cite{Duguet17, Soma20}, in accordance with analogous calculations in quantum chemistry~\cite{vonNiessen1984QCadc,Danovich2011,Degroote2011frpa,Barbieri2012PRA}.
Conservatively, one can associate uncertainty bands with ADC(2)-ADC(3) differences, as done in Fig.~\ref{fig_spectra_CaK} where spectra resulting from one-nucleon addition and removal from and to $^{54}$Ca are displayed. 
These four systems sit at the limits of what can be presently studied in radioactive ion beam facilities. 
Yet, not many experimental data are available, such that these calculations mostly represent predictions that can be presumably tested in the near future.
In $^{53}$Ca, where the ground state and two excited states are identified, \lnl{} GF calculations succeed in reproducing the measured spectrum in with good accuracy. 
\sat{} instead mildly overestimates the splitting between the two excited states.
In $^{53}$K, the difference between the two interactions is even more striking. 
While \lnl{} reproduces the measured first excited state with good precision, \sat{} predicts a wrong ordering of the $3/2^+$ and $1/2^+$ states, resulting in a ground state with incorrect spin.
The potassium chain indeed constitutes an interesting case because of an unusual inversion (at $N=28$) and re-inversion (at $N=32$) of the spins of the ground- and first excited states~\cite{Papuga13, Papuga14}.
As discussed in Refs.~\cite{Sun20, Soma20}, Gorkov GF calculations with the \lnl{} Hamiltonian yield ax excellent reproduction of the whole trend, from $^{37}$K to $^{53}$K.
Such studies, complementary to those focusing on ground-state observables, are not only relevant in relation to current experimental programs, but also provide a unique testing ground for the development of nuclear interactions.

The spectral representation of the one-body GF thus gives access to spectroscopic studies of odd-even nuclei.
If one is interested in the excitations of an even-even system, the two-body GF (in one of its possible time orderings) has to be considered instead.
A typical example is the polarisation propagator, which provides the response of the nuclear system to an external operator.
In analogy to Eq.~\eqref{eq_Lehmann}, its Lehmann representation reads as
\begin{align}
 \Pi_{\gamma \delta, \alpha \beta}(\omega) ~={}& 
 \sum_{n_\pi \neq 0}  \frac{ 
          \bra{\Psi^A_0} a^{\dagger}_{\delta}  	a_{\gamma}   \ket{\Psi^{A}_{n_\pi}}
          \bra{\Psi^{A}_{n_\pi}}  a^{\dagger}_{\alpha} 	a_{\beta}   \ket{\Psi^A_0}
              }{\omega - (E^{A}_{n_\pi} - E^A_0)+ \textrm{i} \eta }  \nonumber\\
 -{}& \sum_{n_\pi \neq 0} \frac{
          \bra{\Psi^A_0}     a^{\dagger}_{\alpha} 	a_{\beta}       \ket{\Psi^{A}_{n_\pi}}
          \bra{\Psi^{A}_{n_\pi}} 	a^{\dagger}_{\delta}  	a_{\gamma}      \ket{\Psi^A_0}
             }{\omega + (E^{A}_{n_\pi} - E^A_0) -\textrm{i} \eta } \; ,
\label{eq_polarisation}
\end{align}
where $n_{\pi}$ label the excited states of the $A$-body system. 
In the numerators, the residues 
\begin{equation}
\mathcal{Z}^{n_{\pi}}_{\alpha \beta} \equiv \bra{\Psi^A_{n_\pi}}  a_{\alpha}^{\dagger} a_{\beta} \ket{\Psi^{A}_0} \, .
\label{Residues}
\end{equation}
represent particle-hole matrix elements between excited states of the $A$-nucleon system.
The poles appearing in the denominator
\begin{equation}
\epsilon_{n_{\pi}}^{\pi} \equiv E_{n_\pi}^{A} -E_0^{A} 
\label{Poles}
\end{equation}
instead constitute energy differences between excited states of the $A$-nucleon system and its ground-state.
The polarisation propagator~\eqref{eq_polarisation} is obtained as a solution of the Bethe-Salpeter equation,
\begin{align}
 \Pi_{\gamma \delta, \alpha \beta}(\omega) ~={}&  \Pi^{(0)}_{\gamma \delta, \alpha \beta}(\omega) \nonumber\\
{}& + \sum_{\mu \rho \nu \sigma} \Pi^{(0)}_{\gamma \delta, \mu \rho}(\omega)  K^{(ph)}_{\mu \rho,  \nu \sigma}(\omega) \Pi_{\nu \sigma, \alpha \beta}(\omega) \; ,
\label{eq:BetheSal}
\end{align}
where $\Pi^{(0)}(\omega) $ is the free polarisation propagator, and the particle-hole irreducible interaction $K^{(ph)}$ plays a similar role as that of the self-energy in Eq.~\eqref{eq_Dyson} for the single-particle propagator.
While the corresponding formalism has been developed and implemented for Dyson GFs, the generalisation to the Gorkov framework remains to be carried out.
\begin{figure}
\centering
\includegraphics[width=8.5cm]{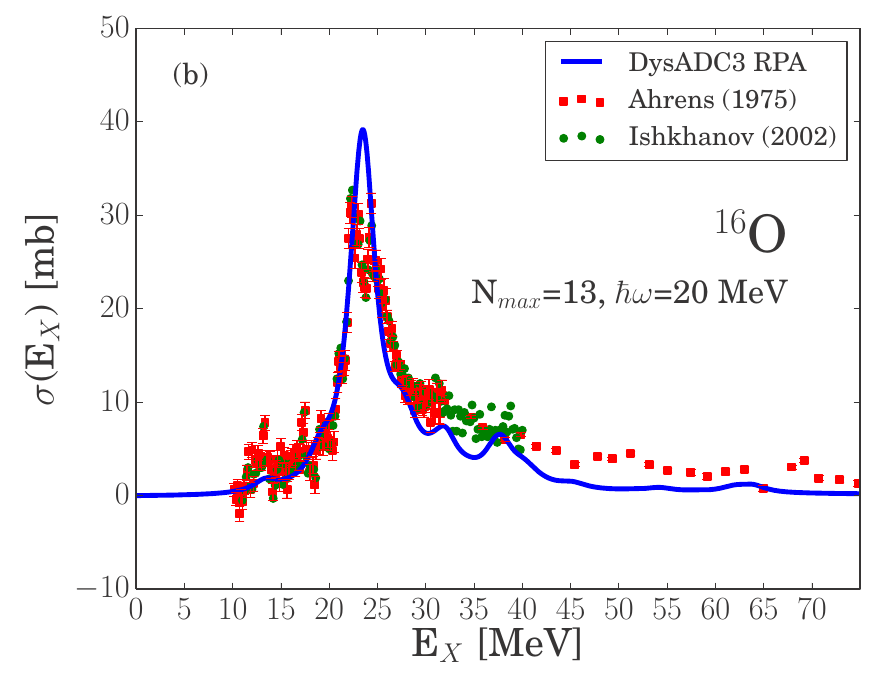}
\quad
\includegraphics[width=8.6cm]{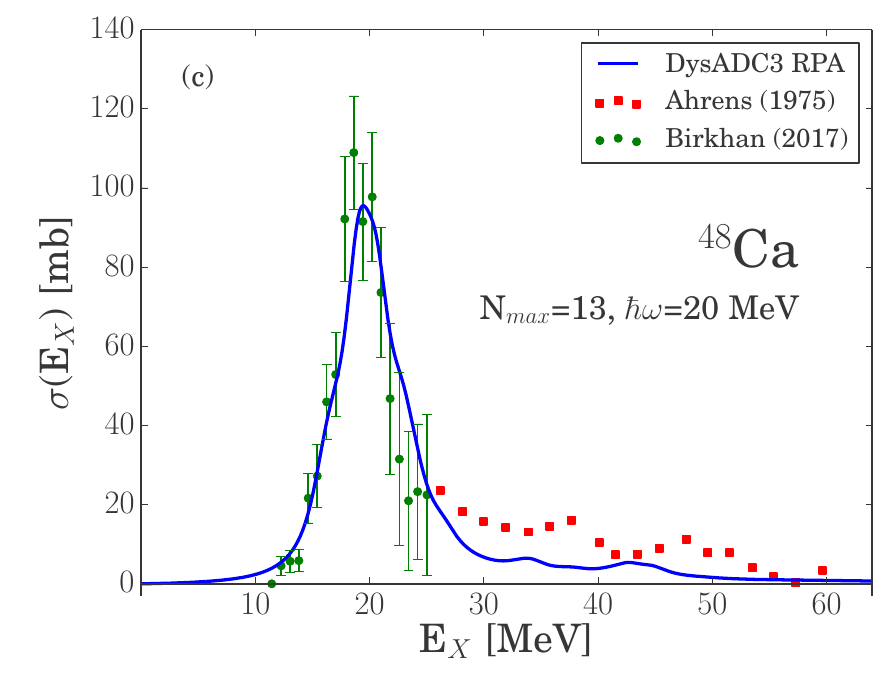}
\caption{
Isovector E1 photoabsorbption cross section of $^{16}$O (left) and $^{48}$Ca (right) computed with the \sat{} interaction.
Theoretical curves are obtained by folding the discrete spectra with Lorentzian widths $\Gamma = 3$ MeV.
Experimental data are from Refs.~\cite{Ahrens75, Ishkhanov02, Birkhan17}.
Adapted figure with permission from Raimondi et al.~\cite{Raimondi19a}, copyright (2019) by the American Physical Society.
}
\label{fig_E1}
\end{figure}

In Refs.~\cite{Barbieri18, Raimondi19a} the polarisation propagator was calculated with the aim of accessing the nuclear isovector electric dipole (E1) response.
A dressed random phase approximation (DRPA) was adopted for the computation of $\Pi(\omega)$.
The scheme makes use of a correlated, e.g. ADC(3), one-body propagator
as the starting point for the RPA equations, thus going beyond a simple HF-based particle-hole resummation.
Figure~\ref{fig_E1} shows results for the integrated isovector E1 photoabsorption cross section, which is directly obtained from the polarisation propagator~\cite{Raimondi19a}.
Two representative examples are shown here, $^{16}$O and $^{48}$Ca.
Calculations make use of the \sat{} interaction, ensuring that nuclei have the correct size, which is a crucial property for this application.
In $^{16}$O the peak associated to the giant dipole resonance (GDR) is well reproduced.
At higher excitation energies, the calculation underestimates the experimental spectrum, presumably due to missing correlations beyond the simple RPA.
A similar picture emerges for $^{48}$Ca, with the GDR peak in good agreement with recent experimental measurements~\cite{Birkhan17} and the high-energy tail missing some strength.
Several other closed-shell nuclei, for some of which experimental data are not yet available, are discussed in Ref.~\cite{Raimondi19a}.
A generalisation of this formalism to the Gorkov framework would allow us to extend these promising results to a large number of open-shell nuclei.

\vskip0.2cm
\subsection{Lepton-nucleus scattering}
\label{sec_lepton}
\vskip0.2cm

The spectral function introduced in Eqs.~\eqref{eq_spectral}-\eqref{eq_sdd} carries information about the energy-momentum distribution of the correlated nucleons.
Certain scattering processes, where an external probe scatters off the nucleus, under certain kinematical conditions (e.g. characterised by a sufficiently large momentum transfer), can be described as an incoherent sum of scattering amplitudes on bound nucleons.
Then, the cross section can be computed in the so-called impulse approximation and expressed as a sum of one-body terms containing a convolution with the nucleon spectral function.

This is the case, e.g., of electron and neutrino scattering in the region of the quasielastic peak.
Here the double differential cross section for inclusive lepton-nucleus scattering can be written as~\cite{PaviaBook}
\begin{align}
\Big(\frac{d\sigma}{dE^\prime d\Omega^\prime}\Big)_\ell &= C_\ell\; \frac{E_k^\prime}{E_k} \; L_{\mu \nu}W^{\mu \nu} \, ,
\label{eq:cross_sec}
\end{align}
where $L_{\mu \nu}$ is the leptonic tensor and  $k=(E_k,{\bf k})$ and $k^\prime=(E_k^\prime , {\bf k}^\prime)$ are the laboratory four-momenta of the incoming and outgoing leptons, respectively.
The factor $C_\ell=\alpha/(k-k^\prime)^4$ for electrons and \hbox{$C_\ell=G/8\pi^2$} for neutrinos, where $G=G_F$ for neutral current (NC) and $G=G_F \cos\theta_c$ for charged current (CC) processes. The electroweak coupling constants are \hbox{$\alpha\simeq 1/137$},  $G_F=1.1803 \times 10^{-5}\,\rm GeV^{-2}$~\cite{Herczeg:1999} and~\hbox{$\cos\theta_c=0.97425$~\cite{Nakamura2010PDG}}. 
The hadron tensor $W^{\mu \nu}$ incorporates the transition matrix elements from the target ground state $|\Psi^A_0\rangle$ to the final states $|\Psi^A_f\rangle$ due to the hadronic currents, including additional axial terms for neutrino scattering.  
The impulse approximation consists in factorising $|\Psi^A_f\rangle\rightarrow|{\bf p}^\prime\rangle \otimes |\Psi^{A-1}_n\rangle$, i.e. it allows to work with the outgoing nucleon of momentum ${\bf p}^\prime$ and the residual nucleus, left in a state $|\Psi^{A-1}_n\rangle$. 
This leads to~\cite{Rocco2018escatt,Rocco20192bcurr}
\begin{align}
&W^{\mu\nu}_{\rm 1b}({\bf q},\omega)=\int \frac{d^3 {\bf p}^\prime \; dE}{(2\pi)^3} \frac{m_N^2}{e({\bf p}^\prime)e({\bf p^\prime\!-\!q})} \delta(\omega+E-e(\mathbf{\bf p}^\prime)) \nonumber \\
&\quad \times  \sum_{s}\,   S^h_s({\bf p}^\prime\!-\!{\bf q},E)  \langle p^\prime | {j_{s}^\mu}^\dagger |p^\prime\!-\!q \rangle \langle p^\prime\!-\!q |  j_{s}^\nu | p^\prime \rangle \, ,
\label{had:tens}
\end{align}
where $\omega$ represents the energy transfer, $M_N$ is the nucleon mass, $e({\bf p})$ the energy of a nucleon with momentum ${\bf p}$.
The one-body current operators ${j}^\mu_s$ depend on the  spin-isospin degrees of freedom $s$ and $S^h_s({\bf p}, E)$ is the one-body spectral function normalised to the total number of nucleons.
For two-body currents and hadron production, Eq.~\eqref{had:tens} extends non trivially in terms of one- and two-body spectral functions~\cite{Giusti2005,Barbieri2004O16epp,Barbieri2006NPB,Rocco20192bcurr}.
\begin{figure}
\centering
\includegraphics[width=8.5cm]{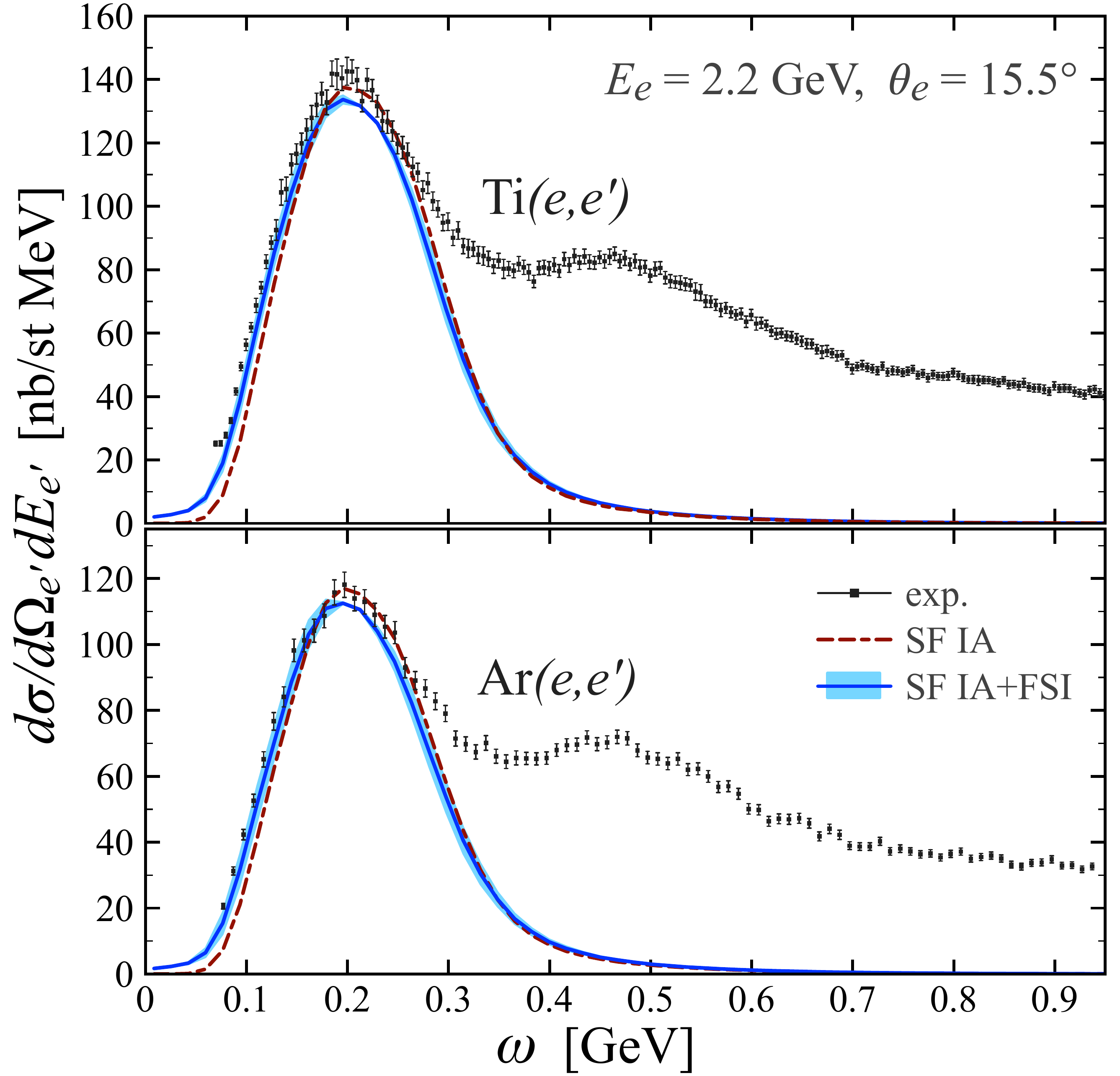}
\quad
\includegraphics[width=8.5cm]{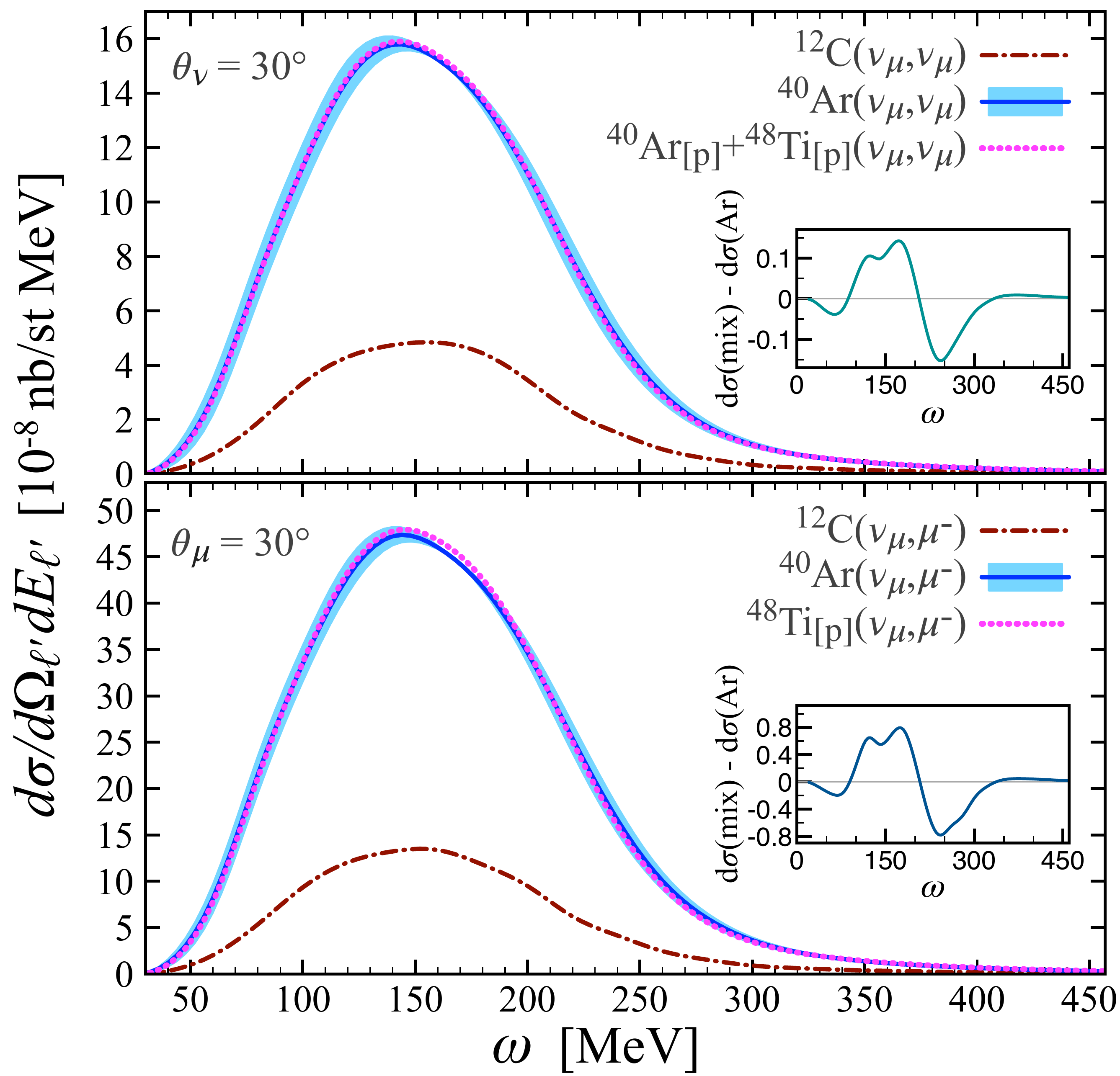}
\caption{
\textit{Left:} 
Inclusive Ti(e, e$^\prime$) (top) and Ar(e, e$^\prime$) (bottom) cross sections at 2.2 GeV and 15.5$^\circ$ scattering angle. 
The solid (dashed) line shows the quasielastic cross section with (without) the inclusion of finals-state interactions (FSI, see Ref.~\cite{Barbieri19} for details). 
For the FSI results, the theoretical uncertainties coming from model-space convergence are also shown as a shaded band. 
Experimental data are taken from Refs.~\cite{Dai2018Tiee1,Dai2019Ar40ee1} and show both the quasielastic peak and the contribution from meson production at larger missing energies.
\textit{Right:} 
Quasielastic neutral (top) and charged current (bottom) cross sections for 1 GeV neutrino scattering. 
Dot-dashed lines refer to a $^{12}$C target and solid lines refer to $^{40}$Ar. 
The coloured band represents theoretical uncertainties due to model-space convergence.
The dotted lines result from using the $^{48}$Ti proton spectral function as an approximation for neutrons in $^{40}$Ar. 
The insets show the difference between the latter and calculations where the full spectral distribution of $^{40}$Ar is used.
Adapted figure with permission from Barbieri et al.~\cite{Barbieri19}, copyright (2019) by the American Physical Society.
}
\label{fig_lepton}
\end{figure}

Following the above formalism, electron and neutrino scattering off argon and titanium isotopes was investigated in Ref.~\cite{Barbieri19}.
The interest in studying lepton scattering off these nuclei resides in the fact that future-generation neutrino experiments (e.g. DUNE~\cite{Acciarri2015dune}) will use liquid-argon time-projection chambers, which rely on scattering of neutrinos off $^{40}$Ar.
The nuclear component of the cross section has to be well determined for a meaningful interpretation of the measured events, in particular to reconstruct the neutrino energy with sufficient accuracy.
In this respect, a tailored electron scattering experiment\footnote{Since the neutron spectral function is not easily accessed by electrons, a complementary study with titanium, which has as many protons (22) as the neutrons in $^{40}$Ar, was carried out.}
was designed and recently performed at JLab~\cite{Benhar14,Dai2018Tiee1,Dai2019Ar40ee1}.
In Ref.~\cite{Barbieri19} nuclear spectral functions of stable argon, calcium and titanium isotopes were computed at the ADC(2) level with the \sat{} interactions.
A good reproduction of available charge radii and density distributions was found.
Starting from these spectral functions, inclusive electron-$^{48}$Ti and electron-$^{40}$Ar cross sections were computed.
They are shown in Fig.~\ref{fig_lepton} (left) as a function of the energy transfer and compared to the recent experimental data from the JLab E12-14-012 collaboration~\cite{Dai2018Tiee1,Dai2019Ar40ee1}.
Calculations closely follow the quasi elastic peak, thus validating the theoretical approach and the impulse approximation in particular.
Next, the quasielastic neutral and charged current cross sections were studied for 1 GeV neutrino scattering.
Results are displayed in the right panel of Fig.~\ref{fig_lepton}, also compared to scattering off $^{12}$C.
The quasielastic peak is found at a similar energy, with an increase in the magnitude of the cross section consistent with super scaling properties of inclusive reactions.
The use of a Ti \textit{proton} spectral function instead of an Ar \textit{neutron} spectral function was also tested and found to be an excellent approximation.
Further studies will be needed to thoroughly assess theoretical uncertainties.

\vskip0.2cm
\section{Perspectives and challenges}
\label{sec_perspectives}
\vskip0.2cm

\vskip0.2cm
\subsection{Towards the description of nuclear reactions}
\label{sec_reactions}
\vskip0.2cm

While the good knowledge of electro(weak) interactions facilitate the modelling of lepton-nucleus scattering, nuclear reactions, because of the complexity of strong interactions, require a more involved theoretical description.
As a consequence, very few ab initio methods are presently capable of going beyond structure properties and directly computing reaction observables.
Nevertheless, nucleon-nucleus and nucleus-nucleus reactions constitute nowadays the tools of choice to study the properties of atomic nuclei, such that progress on their ab initio description would be highly valuable.

\begin{figure}
\centering
\includegraphics[width=6.5cm]{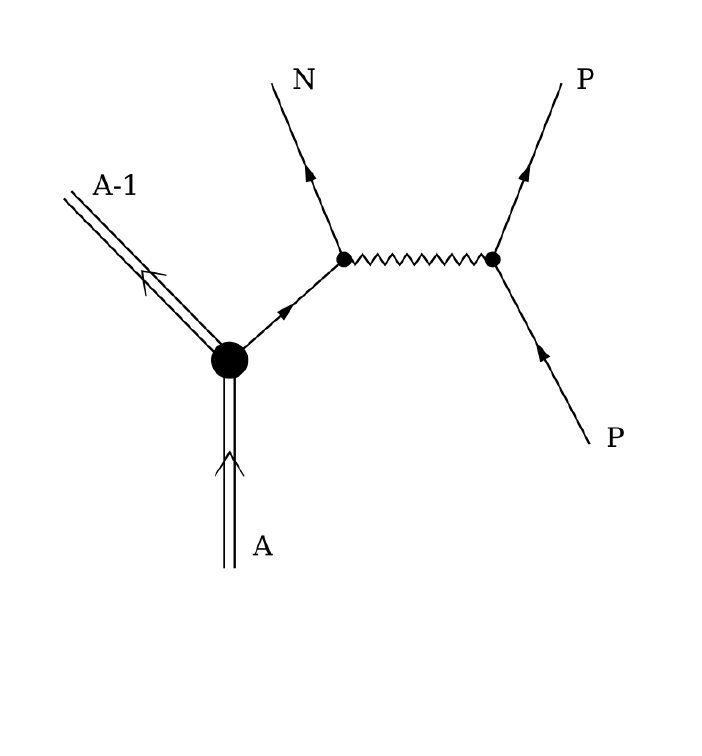}
\quad
\quad
\quad
\includegraphics[width=9cm]{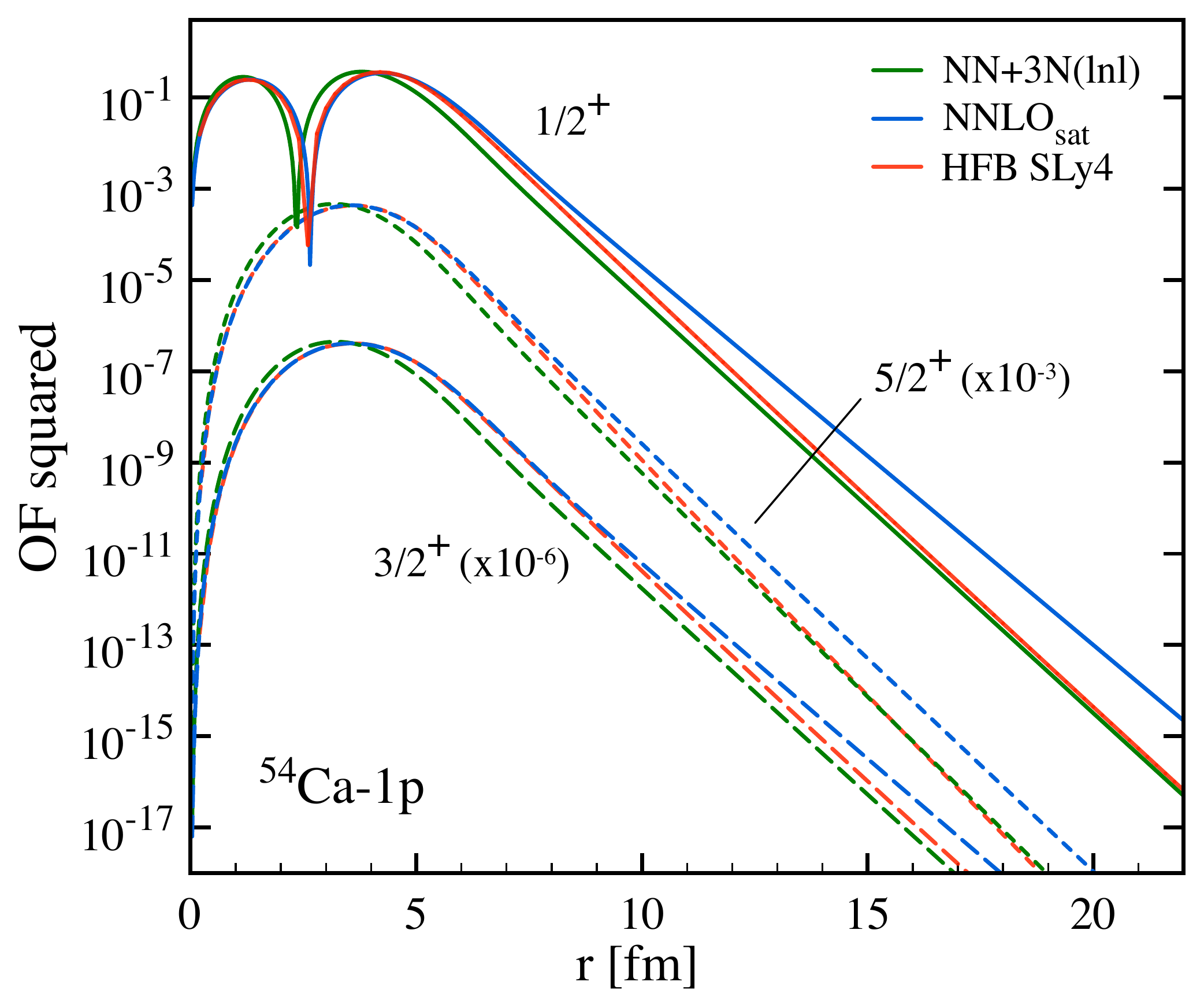}
\caption{
\textit{Left:} 
Schematic representation of quasifree nucleon knockout: an incoming proton $p$ scatters off an $A$-body nucleus, knocking out a nucleon $N$.
\textit{Right:} 
One-proton overlap functions (squared) computed at the ADC(3) level using \sat{} and \lnl{} interactions~\cite{Raimondi19c}.
OFs corresponding to the main quasiparticle fragments for three angular momentum and parity channels $J^\Pi = 1/2^+, 3/2^+, 5/2^+$ are shown.
For comparison, EDF calculations~\cite{bennaceur05a, Sun19} performed with the SLy4 parameterisation are displayed.
These OFs were employed in the calculation of one-proton knockout cross sections in Ref.~\cite{Sun20}.
}
\label{fig_of}
\end{figure}

A relatively clean process that can be used as a training field for ab initio reaction theory is quasifree nucleon knockout~\cite{Aumann13}.
Quasifree scattering represents a process in which an incident nucleon (typically a proton\footnote{Since many current experiments concern unstable nuclei, they are performed in inverse kinematics, whence the use of a proton target. In the following, the case of an incident proton will be thus considered.}) with an energy of few hundred MeV knocks out a bound nucleon from the isotope of interest (see Fig.~\ref{fig_of}, left panel, for a schematic illustration).
Kinematical conditions are chosen such that the process can be preferentially described by a single, localised interaction between the incident and the struck nucleons, thus minimising multiple collisions for the incoming nucleon.
This sudden removal mechanism suggests that the remaining $A-1$ nucleons can be treated as spectators, which translates into an impulse approximation analogous to the one discussed in Sec.~\ref{sec_lepton}.
As a result, the total cross section can be separated into a structure and a reaction part. 
The latter involves nucleons that are usually described in terms of distorted waves to account fo their propagation under the influence of the nuclear medium.
Under these assumptions the differential cross section, labelled distorted-wave impulse approximation (DWIA), can be schematically written as
\begin{equation}
\left ( \frac{d \sigma}{d^3Q} \right )_{\text{DWIA}} =
SF_N \times  \left \langle \frac{d \sigma_{pN}}{d \Omega} \right \rangle
\times
\left \langle 
S_{pA}
S_{p(A-1)}
S_{N(A-1)}
\Psi_N
\right \rangle \; .
\label{eq_qf}
\end{equation}
The first factor on the right-hand side is the spectroscopic factor of the struck nucleon, encoding the structure properties of the nucleus.
The second term represents an in-medium proton-nucleon cross section, determining the probability of the collision between the projectile and one of the bound nucleons.
The third term contains the scattering matrices for the (effective) degrees of freedom at play, i.e. respectively i) proton and initial $A$-body system, ii) proton and final $(A-1)$-body system, iii) struck nucleon and final $(A-1)$-body system, plus the scattering wave function of the outgoing nucleon.

An ab initio calculations for some of the above quantities was recently performed within the GF formalism, in connection with a quasifree neutron knockout study on $^{54}$Ca~\cite{Chen19}.
Specifically, spectroscopic factors and one-body overlap functions for the struck neutron were computed.
The latter enter the evaluation of the scattering matrices, where they are convoluted with nucleon-nucleus optical potentials.
Overlaps functions (OFs) for three different neutron states, corresponding to the first three (ground and excited) states in $^{53}$Ca, are shown in Fig.~\ref{fig_of} (right).
GF calculations were performed at the ADC(3) level with two different interactions, \sat{} and \lnl.
For comparison, OFs from an EDF calculation with the SLy4 Skyrme parameterisation are displayed.
An overall good agreement is found between the three sets of calculations.
Looking more in detail, one notices that \sat{} better reproduces the maximum of the overlap at around 2-5 fm, i.e. the region of the nuclear surface. 
This clearly relates to the ability of this interaction to well describe nuclear sizes (cf. discussion in Sec.~\ref{sec_gs}).
In contrast, \lnl{} is in better agreement in the tail of the OF.
This is in line with the more accurate description of the low-lying spectrum of $^{53}$Ca, cf. Fig.~\ref{fig_spectra_CaK}.

In order to compute the cross section~\eqref{eq_qf} for the $^{54}$Ca($p,pn$)$^{53}$Ca reaction, in Ref.~\cite{Chen19} the GF input was complemented by phenomenological optical potentials and in-medium nucleon-nucleon cross sections, yielding results in good agreement with shell model calculations.
An analogous study had been performed previously for the $^{14}$O($d,t$)$^{13}$O and $^{14}$O($d,^3He$)$^{13}$N reactions, leading to similar conclusions~\cite{Flavigny13}.
These applications might be seen as a first step towards a consistent approach to structure and reaction, and show that ab initio ingredients can be as efficient as phenomenological ones.
In fact, the in-medium nucleon-nucleon cross section could be already extracted from nuclear matter calculations, see e.g. Ref.~\cite{Rios12}.
Moreover, a nucleon-nucleus potential can be directly computed from the one-body self-energy, see Ref.~\cite{Idini19} for the first applications to oxygen and calcium isotopes.
Thus, in the future the full ab initio calculation of the cross section~\eqref{eq_qf} can be envisaged.

\vskip0.2cm
\subsection{Towards doubly open-shell nuclei}
\label{sec_doubly}
\vskip0.2cm

The development of Gorkov GF theory~\cite{Soma11}, and its subsequent implementation to finite nuclei~\cite{Soma13, Soma14a} proved that symmetry breaking can be a powerful tool in the context of ab initio calculations.
The generalisation of other many-body techniques to a symmetry-breaking scheme~\cite{Hergert13, Signoracci15, Tichai18} further confirmed the validity of this strategy.
Such advances have allowed to extend the reach of ab initio calculations in mid-mass systems from a few closed-shell nuclei to a large number of open-shell isotopes, e.g. to complete semi-magic isotopic or isotonic chains.
In their current implementation, however, these methods do not break rotational symmetry.
This results in an inefficient account of quadrupole correlations, such that the description of (doubly) open-shell systems displaying significant deformation can be problematic.

As opposed to pairing, where the strong static correlations at the Fermi surface cause the breakdown of the particle-hole expansion, in the presence of deformation one can usually produce converged calculations, i.e. compute few orders in the many-body expansion.
Nevertheless, one expects the accuracy to deteriorate with the strength of the deformation. 
This has been indeed observed in Gorkov GF calculations around the calcium chain, in particular for titanium and chromium isotopes characterised by mid-shell protons.
For instance, by studying neutron gaps\footnote{Neutron gaps, defined as differences of two-neutron separation energies, are one of possible observables that meaningfully estimate the ''magic'' character of a neutron number.} at the neutron traditional magic number $N=28$ one finds an excellent agreement with experiment up to $Z=22$, after which symmetry-restricted GF calculations clearly depart from data~\cite{Mougeot20}.
Furthermore, one can identify a correlation between the deviation to experimental data and the amount of deformation (e.g. estimated by EDF calculations~\cite{Bender06}).
This situation suggests that the additional breaking of the SU(2) symmetry associated to rotational invariance will be needed in the extension of correlation-expansion methods to doubly open-shell nuclei. 
Some of the existing approaches are indeed being generalised along these lines~\cite{Yao19}.
In the case of Gorkov GF such an extension will presumably require the use of importance truncation~\cite{Tichai19b} and/or tensor factorisation techniques~\cite{Tichai19a}.

\vskip0.2cm
\subsection{Towards heavy nuclei}
\label{sec_heavy}
\vskip0.2cm

Over the past years, GF ab initio calculations have extended their reach across the Segr\`e chart, going from the first application to the oxygen chain ($A\sim20$)~\cite{Cipollone13} to recent computations of nickel isotopes ($A\sim70$)~\cite{Soma20}.
Such calculations rely on sophisticated numerical codes that make extensive use of available high-performance computing resources.
Although the management of the computing time\footnote{In this respect, while the numerical cost of Dyson GF calculations grows with the mass number, the one of Gorkov GFs solely depend on the basis dimension~\cite{Soma14a}.} could be problematic, the bottleneck that currently prevents (converged) calculations beyond $A\sim100$ is related to the storage of the matrix elements of $3N$ operators.
Indeed, presently employed truncations on the three-body basis of $e_{3\text{max}}=14-18$ allow keeping the size of $3N$ matrix elements below 100 GB, which is roughly the order of magnitude of the available memory on a single node of a state-of-the-art supercomputer.
Going considerably beyond this size thus constitutes an issue.
At the same time, such values of $e_{3\text{max}}$ are enough to achieve reasonably converged results in the ($A\sim60-70$), while they become insufficient for larger isotopes~\cite{Hagen16, Soma20}.

Different strategies are being explored to overcome this issue. 
One possibility would be to discard beforehand, for a given $e_{3\text{max}}$, a subset of the initial $3N$ matrix elements.
While performing a selection\footnote{I.e., using a different truncation than the $e_{3\text{max}}$ truncation.} directly on the original set might be problematic, a promising technique based on tensor factorisation algorithms has been put forward recently~\cite{Tichai19b}.
Since in the majority of applications $3N$ forces are included in the normal-ordered two-body approximation\footnote{It consists of two steps: i) a normal ordering of the ($3N$) Hamiltonian with respect to a reference state and ii) the disregard of operators of rank higher than two (see Ref.~\cite{Ripoche20} for a pedagogical description and the generalisation to the case of symmetry-breaking reference states).}, another option could consist in performing the normal ordering procedure in a different (smaller) basis than the HO one, e.g. in momentum space.

\begin{figure}
\centering
\includegraphics[width=8.52cm]{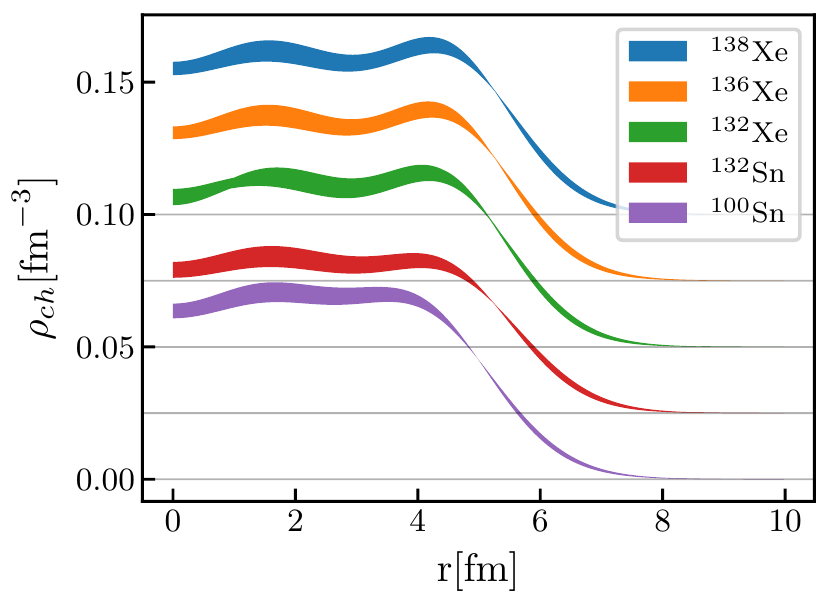}
\hfill
\includegraphics[width=8.5cm]{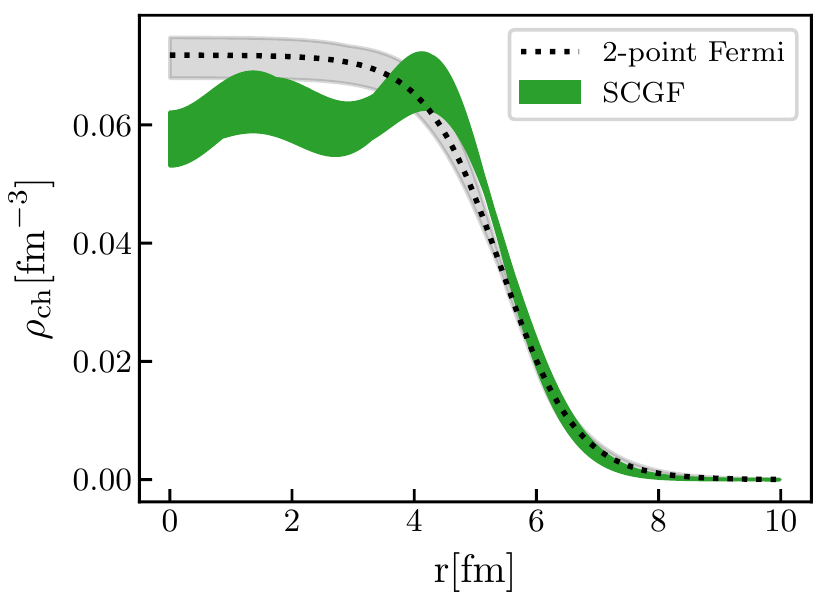}
\caption{
\textit{Left:} 
Charge density distributions of selected Sn and Xe isotopes computed with the \sat{} interaction in Gorkov GF theory at the ADC(2) level.
Each curve is shifted upwards with respect to the one below by 0.025 fm$^{-3}$ for clarity purposes.
Coloured band account for uncertainties stemming from model-space convergence.
\textit{Right:} 
Charge density distributions of $^{132}$Xe computed with the \sat{} interaction in Gorkov GF theory at the ADC(2) level.
The dotted line with grey band corresponds to the two-point Fermi distribution with parameter and error bars extracted from Ref.~\cite{Tsukada17}.
Adapted figure with permission from Arthuis et al.~\cite{Arthuis20}.
}
\label{fig_xe}
\end{figure}
Even with current limitations, however, some meaningful results can be produced for nuclei above mass $A=100$.
Indeed, as discussed in Sec.~\ref{sec_basis}, not all observables show the same convergence pattern.
In particular, while ground-state energy curves get lower and lower as the basis size is increased, radii tilt around a fixed point that can be assumed to correspond to the infinite-basis result.
This allow to provide an estimate of the radius (and, correspondingly, of the density distribution) in bases for which the energy is far from being converged.
Based on this observation, charge densities of closed- and open-shell tin and xenon isotopes have been recently calculated within GF theory~\cite{Arthuis20}.
Examples are reported in Fig.~\ref{fig_xe}, where the \sat{} interaction has been employed.
In the left panel, the charge distribution of several nuclei is shown, exemplifying the typical range of system that can be presently accessed.
In the right panel, the charge density of $^{132}$Xe is displayed and compared to a two-point Fermi distribution fitted on the recent experimental data from the SCRIT collaboration~\cite{Tsukada17}.
The two are in very good agreement at the surface and in the tail of the distribution.
In the interior, the two-point Fermi distribution and the lack of high-momentum transfer data lead to flat behaviour for the experimental distribution, whereas the computed density shows a well defined oscillation pattern.
This example shows that even in present implementations of GF calculations it is possible to provide relevant predictions in the region $A=100-150$.

\vskip0.2cm
\section*{Acknowledgements}
\vskip0.2cm

The author wishes to thank F. Raimondi and Y. L. Sun for providing the results appearing in Fig. 9, all collaborators that contributed to the results presented in this work, and additionally C. Barbieri and T. Duguet for useful remarks on the manuscript.

\bibliographystyle{frontiersinHLTH_FPHY} 
\bibliography{biblio_frontiers}

\end{document}